%% file: main.tex
\documentclass[fleqn,usenatbib]{mnras}
\usepackage{times}
\usepackage{latexsym}

\input Planck.tex

\usepackage{microtype}

\usepackage{caption}
\usepackage{subcaption}
\usepackage{amssymb}
\usepackage{afterpage}
\usepackage{float}
\usepackage{caption}
\usepackage[section]{placeins}
\usepackage{multirow}
\usepackage{verbatim}
\usepackage{amsmath}
\usepackage{textpos}
\usepackage{ragged2e}
\usepackage{array}
\usepackage{commath}
\usepackage{fancyhdr}
\usepackage{fancyvrb}
\usepackage{graphicx}
\usepackage{braket}
\usepackage{bm}
\usepackage{wrapfig}

\graphicspath{ {./Images/} }

\def\fnl{f_{\mathrm{NL}}}
\def\fnlloc{f_{\mathrm{NL}}^{\text{local}}}
\def\fnlequi{f_{\mathrm{NL}}^{\text{equil}}}
\def\fnlortho{f_{\mathrm{NL}}^{\text{orth}}}
\def\fnliswlens{f_{\mathrm{NL}}^{\text{ISW-lens}}}
\def\tripalm{a_{\ell_1m_1}a_{\ell_2m_2}a_{\ell_3m_3}}
\mathcode`l="8000
\begingroup
\makeatletter
\lccode`\~=`\l
\DeclareMathSymbol{\lsb@l}{\mathalpha}{letters}{`l}
\lowercase{\gdef~{\ifnum\the\mathgroup=\m@ne \ell \else \lsb@l \fi}}%
\endgroup

\title{Biases to primordial non-Gaussianity measurements from CMB secondary anisotropies}

\author[Coulton, Miranthis and Challinor]{William Coulton$^{1}$, Alexander Miranthis$^{2}$,  Anthony Challinor$^{2,3}$ \\$^{1}$Center for Computational Astrophysics, Flatiron Institute, New York City, NY, USA
\\$^{2}$DAMTP, Centre for Mathematical Sciences, Wilberforce Road, Cambridge CB3 0WA, UK
\\$^{3}$Institute of Astronomy and Kavli Institute for Cosmology, Madingley Road, Cambridge, CB3 0HA, UK}

\date{}

\begin{document}
\maketitle

\begin{abstract}
Our view of the last-scattering surface in the cosmic microwave background (CMB) is obscured by secondary anisotropies, sourced by scattering, extragalactic emission and gravitational processes between recombination and observation. Whilst it is established that non-Gaussianity from the correlation between the integrated-Sachs--Wolfe (ISW) effect and gravitational lensing can significantly bias primordial non-Gaussianity (PNG) searches, recent work by Hill (2018) has suggested that other combinations of secondary anisotropies can also produce significant biases. Building on that work, we use the WebSky and Sehgal et al.~(2010) simulations to perform an extensive examination of possible biases to PNG measurements for the local, equilateral and orthogonal shapes.
For a \textit{Planck}-like CMB experiment, without foreground cleaning, we find significant biases from 
cosmic infrared background (CIB)-lensing and thermal Sunyaev--Zel'dovich (tSZ)-lensing bispectra for the local and orthogonal templates, and from CIB and tSZ bispectra for the equilateral template. For future experiments, such as the Simons Observatory, biases from correlations between the ISW effect and the tSZ and CIB will also become important.   Finally we investigate whether foreground-cleaning techniques are able to suppress these biases sufficiently. We find that the majority of these biases are effectively suppressed by the internal-linear-combination method and the total bias to \textit{Planck}-like and SO-like experiments is less than the $1\,\sigma$ statistical error.
However, the small total bias arises from the cancellation of several $1\,\sigma$ biases for \textit{Planck}-like experiments and $2\,\sigma$ biases for SO-like. As this cancellation is likely sensitive to the precise modelling, to ensure robustness against these biases explicit removal methods should be used, likely at the cost of decreased constraining power. 
\end{abstract}

\section{Introduction}
The cosmic microwave background (CMB) is the relic radiation from the big bang and provides a snapshot of the universe at the time of recombination, around $380,000$ years after the big bang. The radiation is highly consistent with blackbody radiation with a temperature of  $T_{\mathrm{CMB}}=(2.7255 \pm 0.0006)\,\text{K}$~\citep{fixsen2009}. After subtracting the dipole due to the Earth's motion \citep{Conklin_1969,Henry_1971,planck2013-pipaberration}, there are anisotropies in the CMB temperature of order $10^{-5}\,\text{K}$~\citep{Smoot_1992,Bennett_2003}. These are directly related to primordial fluctuations and are key to placing constraints on and improving our understanding of inflation and the origins of our universe \citep{SZ_1970,Peebles_1970}. Through an array of CMB experiments the primordial perturbations have been found to be statistically isotropic, highly Gaussian and close to scale invariant \citep{WMAP9_cosmology,planck2016-l01,ACT_2012_Power,ACT_SPT_WMAP_NG}. Whilst these measurements have transformed our understanding of the early universe, there still remains a very broad range of models that are consistent with these observations \citep{Linde_2014,Brandenberger_2015,Ijjas_2016}. 

One promising avenue to distinguish between models of the early universe is primordial non-Gaussianity (PNG). Primordial non-Gaussianity characterizes how the perturbations in the early universe deviate from a Gaussian distribution.  Many models of the universe produce large deviations from non-Gaussianity  with a rich phenomenology \citep[see, e.g.,][]{Chen_2010,Lehners_2010} that can be used to infer details of the early universe such as the field content or strength of interactions.  In this work we will focus on deviations from Gaussianity that produce a non-zero three-point function, or its harmonic equivalent, the bispectrum, as for many models this is the strongest statistical probe. There has been extensive work developing methods to constrain PNG optimally \citep{Komatsu_2005,Smith_2011,Fergusson_2010,Bucher_2010} and these have been applied to many data sets \citep[e.g.,][]{Komatsu_2001,Senatore2010}.  Whilst the current best limits, coming from the \textit{Planck} satellite   \citep{planck2016-l09}, already impose stringent constraints on the physics of the early universe, future constraints  have the possibility to rule out large classes of early-universe models and a detection would be transformative \citep{meerburg2019primordial}.

An inherent difficulty in making observations of the primary CMB temperature anisotropies is the presence of both Galactic emissions and CMB secondary anisotropies: temperature fluctuations imprinted onto the CMB as the photons propagate through the universe and confounding emission from extragalactic sources \citep{Aghanim_2008}. 
Secondary anisotropies arise from a range of effects including: the integrated Sachs--Wolfe (ISW) effect \citep{Sachs_1967,ISW_Rees-Sciama_1968}; the thermal and kinetic Sunyaev--Zeldovich (tSZ and kSZ) effects \citep{SZ_1980_tSZeffect,Sunyaev_1972}; gravitational lensing \citep{Blanchard1987,Zaldarriaga_1999,Bernardeau1997}; radio galaxies \citep{Fanaroff_1974,Lin_2009}; and the cosmic infrared background (CIB) \citep{Lagache2007,Viero_2009}. Whilst the impact of Galactic emissions on primordial non-Gaussianity studies has been studied in detail \citep{Jung_2018}, an exhaustive study of the impact of CMB secondaries has yet to be performed and is the focus of this work.

These secondary anisotropies can bias measurements of PNG as they are non-Gaussian and correlated with each other \citep{Crawford_2014,planck2013-p05b,planck2013-pip56,Coulton_2018}. Most work to date has focused on removing the most significant effect of these, the ISW-lensing bispectrum, which peaks strongly in the shape of local-type PNG \citep{Lewis_2011,Smith_2011,Serra_2008,Mangilli_2009,Hanson_2009}. For the \textit{Planck} experiment, the bias that had to be removed was $\Delta\fnlloc{}^{\text{,ISW-}\phi}=7.5$ \citep{planck2014-a19}. However, recent work by \citet{Hill_2018,Curto_2015} has indicated that there are potentially significant biases from other foreground bispectra that are currently unaccounted for.

One avenue for mitigating these effects is utilizing multi-frequency observations to separate out the CMB secondaries from the primary CMB \citep[see, e.g.,][for a review]{Ichiki2014}. The tSZ effect, radio galaxies and the CIB do not follow a blackbody distribution, and thus by combining maps of the CMB at different frequencies one can attempt to isolate these effects. The other three sources of secondary anisotropies (ISW, lensing and kSZ), however, do not alter the blackbody spectrum of the CMB and so cannot be mitigated with such techniques. Any biases arising solely from these sources must be subtracted off (as was done with the ISW-lensing bispectrum for the results of \textit{Planck}). To date there has not been a thorough exploration of how effective component-separation methods are for suppressing the PNG biases from CMB secondaries.  We note that the CMB temperature anisotropies are also obscured by Galactic emissions that are non-Gaussian, however multi-frequency cleaning has been shown to be sufficiently effective at removing these \citep{Jung_2018,Coulton_2019}.

In this work, we measure the bispectra of these secondary anisotropies directly from the WebSky \citep{Stein_2018,Stein_2020} and \citet{sehgal_2010} simulations. These are a set of simulated full-sky maps of each of the secondary anisotropies. As each is map is generated from the same cosmological realisation of the underlying large-scale structure (LSS), they importantly contain the correct correlations between the secondary anisotropies. The WebSky suite was run at a set of cosmological parameters consistent with the observations by the \textit{Planck} satellite \citep{planck2016-l06} and the secondary anisotropies were tuned to match their power spectra as observed by \textit{Planck} \citep{planck2014-a28,planck2014-a29,planck2013-pip56,planck2014-a26,planck2016-l08}. The Sehgal et al.\ simulations were run with cosmological parameters fit from WMAP-5 \citep{Komatsu2009}, but we use the updated maps from \citet{SO_2019} that are improved to match better the observations of \textit{Planck}. With both sets of simulations at hand, we are able to assess somewhat the theoretical uncertainties present in our results.

We use the binned bispectrum estimator, as developed by \citet{Bucher_2010,BinnedEST_2016}, to measure these bispectra and then project these onto primordial templates to assess how they bias PNG measurements. This approach differs from past work \citep[e.g.,][]{Hill_2018}, where calculations were done analytically based on modelling of the different secondary anisotropies. Our approach has the benefit of easily being extended to many further bispectra, which are not easily or accurately computed analytically, and also by being able to be applied straightforwardly to component-separated maps to evaluate their effectiveness.  We cross-check our methods against analytical results, where they exist, to further validate our approach.

This paper is organised as follows. In Sections \ref{sec:backgroundContext} and \ref{sec:bispectrumEst}  we further motivate the study of PNG and outline the methodology of our approach and the binned estimator. In Section \ref{sec:secondaryAnisotropies} we briefly review the secondary anisotropies considered in this work and Section \ref{sec:simulations} describe the WebSky and Sehgal et al.\ simulations used in this work. In Section \ref{sec:resultsFreqChan} we present the results of using the estimator on single frequency observations of the  WebSky and Sehgal et al.\ maps , comparing these values to those calculated in \citet{Hill_2018}. In Section \ref{sec:CleanedResults} we explore how well these biases are suppressed by multi-frequency component-separation methods estimates for \textit{Planck}-like and Simons Observatory-like experiments \citep{SO_2019}. We conclude in Section \ref{sec:conclusion}.

\section{Primordial non-Gaussianity and the CMB bispectrum} \label{sec:backgroundContext}

Within an inflationary model, we can compute the statistics of the primordial curvature perturbation, $\zeta(\bm{x},t)$, which physically represents how the local scale factor of the universe differs from that of an unperturbed, FRW universe: $a(\bm{x},t)=a_{\mathrm{hom}}(t)e^{\zeta(\bm{x},t)}$ \citep{Bardeen_1980,Kodama_1984,Mukhanov_1992,Wands_2000}. 

The primordial bispectrum, $B_\zeta(k_1,k_2,k_3)$, is the Fourier equivalent of the three-point correlation function and is the lowest-order non-Gaussianity statistic. It is defined as
\begin{equation}
\braket{\zeta(\bm{k}_1)\zeta(\bm{k}_2)\zeta(\bm{k}_3)}=(2\pi)^3\delta^{(3)}(\bm{k}_1+ \bm{k}_2 +\bm{k}_3)B_{\zeta}(k_1,k_2,k_3),
\end{equation} 
where, under the assumptions of statistical isotropy and translational invariance, the bispectrum depends only on the lengths of the wavenumbers ($k_i=|\bm{k}_i|$) and these are enforced to form a closed triangle.

The structure of the primordial bispectrum encodes the rich physics of the primordial universe \citep[see, e.g.,][for reviews]{Chen_2010, meerburg2019primordial}. In this work we focus on three commonly studied configurations, also known as shapes and templates, of the bispectrum. First, the local bispectrum 
\begin{multline}
 B^{\rm local}_{\zeta}(k_1,k_2,k_3) =  \dfrac{6}{5} f^{\rm local}_{\mathrm{NL}}A_s^2k_p^{8-2n_s} \left[ \frac{1}{k_1^{4-n_s}k_2^{4-n_s}} \right. \\
 \left. + \frac{1}{k_2^{4-n_s}k_3^{4-n_s}}+\frac{1}{k_1^{4-n_s}k_3^{4-n_s}} \right],
\end{multline}
where $n_s$ is the spectral index of the power spectrum of $\zeta$, $A_s$ is the amplitude of the power spectrum at pivot scale
$k_p$ and $f^{\rm local}_{\mathrm{NL}}$ is the bispectrum amplitude. This bispectrum is of great interest as it is a powerful probe of the field content of the early universe; \citet{Maldacena_2003} and \citet{Creminelli_2004} showed that within all single-field inflationary models the local bispectrum amplitude is proportional to the slow-roll parameters, hence is highly suppressed. In multi-field inflationary models, however, local-type non-Gaussianity can be large \citep[see, e.g.,][]{Sasaki_2006} and thus a detection of this type of non-Gaussianity would be highly informative.
Second, we consider the equilateral bispectrum
\begin{multline}
B^{\rm{equil}}_{\zeta}(k_1,k_2,k_3) = \dfrac{18}{5} A_s^2 k_p^{8-2n_s} f^{\rm {equil}}_{\mathrm{NL}}\left(-\frac{1}{k_1^{4-n_s}k_2^{4-n_s}}  \right.\\ \left.  - \frac{1}{k_2^{4-n_s}k_3^{4-n_s}}-\frac{1}{k_1^{4-n_s}k_3^{4-n_s} }-\frac{2}{(k_1k_2k_3)^{2(4-n_s)/3} } \right.  \\  \left.+\left[\frac{1}{k_1^{(4-n_s)/3}k_2^{(4-n_s)/3}k_3^{2(4-n_s)/3}} +\text{5 perm.}\right] \right),
\end{multline}
and, finally, the orthogonal bispectrum
\begin{multline}
B^{\rm{orth}}_{\zeta}(k_1,k_2,k_3) = \dfrac{18}{5} A_s^2 k_p^{8-2n_s} f^{\rm {orth}}_{\mathrm{NL}}\left(-\frac{3}{k_1^{4-n_s}k_2^{4-n_s}} \right. \\ \left. - \frac{3}{k_2^{4-n_s}k_3^{4-n_s}} -\frac{3}{k_1^{4-n_s}k_3^{4-n_s} }-\frac{8}{(k_1k_2k_3)^{2(4-n_s)/3} } \right.  \\ \left.+\left[ \frac{3}{k_1^{(4-n_s)/3}k_2^{(4-n_s)/3}k_3^{2(4-n_s)/3}} +\text{5 perm.}\right] \right),
\end{multline}
which are both probes of  non-linear  dynamics during inflation \citep{Creminelli2006,Senatore2010}. \citet{Baumann_2016} and \citet{Green_2020} found that equilateral and orthogonal non-Gaussianity can only be large if the sound speed of the primordial fluctuations is small.

In general, the curvature perturbations generated in inflation are related to the temperature anisotropies of the CMB via the radiation transfer functions, $T_{\ell}(k,\eta_0)$. These are the line-of-sight solution to the linearized Boltzmann equations \citep{Seljak_1996} and we compute these using the public Boltzmann code CAMB \citep{Lewis:1999bs}. The spherical harmonic coefficients of the temperature anisotropies, at present conformal time $\eta_0$ and position $\bm{x}_0$, are then given by 
\begin{align}
a_{\ell m}=4\pi i^{\ell}\int\frac{d^3\bm{k}}{(2\pi)^3}\zeta(\bm{k})T_{\ell}(k,\eta_0)Y^*_{\ell m}(\hat{\bm{k}}),
\end{align}
where $a_{\ell m}$ are the spherical harmonics of the CMB temperature anisotropies:
\begin{align}
\frac{\Delta T}{T}=\sum_{\ell, m}a_{\ell m}Y_{\ell m}(\hat{\bm{n}}).
\end{align}
 In Section \ref{sec:bias_intuition} we will examine the large-scale temperature anisotropies, where the transfer function is trivial $T_\ell(k,\eta_0) = -1/5$. This limit is know as the Sachs--Wolfe limit \citep{Sachs_1967} and is useful for building intuition about possible biases. 

The linear nature of the transfer functions means that we can probe the primordial bispectrum through studying the CMB bispectrum by \citep{Spergel_1999}
\begin{align}
\braket{\tripalm}=b_{\ell_1\ell_2\ell_3}\mathcal{G}^{m_1m_2m_3}_{\ell_1\ell_2\ell_3},
\label{eq:reducedbispectrum}
\end{align}
where we define the reduced bispectrum, $b_{\ell_1\ell_2\ell_3}$, and  $\mathcal{G}^{m_1m_2m_3}_{\ell_1\ell_2\ell_3}$ is the Gaunt integral
\begin{align}
\mathcal{G}^{m_1m_2m_3}_{\ell_1\ell_2\ell_3}=\int d^2\hat{\bm{n}}Y_{\ell_1m_1}(\hat{\bm{n}})Y_{\ell_2m_2}(\hat{\bm{n}})Y_{\ell_3m_3}(\hat{\bm{n}}).
\end{align}
The Gaunt integral arises due to the rotational invariance (or equivalently angular momentum conservation) of the bispectrum and imposes a set of geometrical constraints. The physical information is thus contained in the reduced bispectrum and is related to the primordial bispectrum as \citep{Komatsu_2001}
\begin{align}
b_{\ell_1,\ell_2,\ell_3}  = \int r^2 \mathrm{d}r \prod \limits_i & \left[\int 2\frac{\mathrm{d}k_i}{\pi} T_{\ell_i}(k_,\eta_0) j_{\ell_i}(k_i r,\eta_0) \right]\nonumber\\
& \times B_\zeta(k_1,k_2,k_3),
 \end{align}
where
$j_{\ell}(kr)$ are the spherical Bessel functions, arising from projecting the 3D anisotropies onto the 2D spherical surface of last scattering.

\section{ Bispectrum estimation}\label{sec:bispectrumEst}
Estimating the full CMB bispectrum from data maps is computationally prohibitive and so a range of methods have been developed to study primordial non-Gausianity efficiently~\citep{Komatsu_2005,Fergusson_2010,Bucher_2010}. In this work we fit templates to a coarse-grained measurement of the bispectrum and describe the details of these methods in this section.
\subsection{Binned bispectrum estimator}
 In this paper we implement a binned estimator as developed in \citet{Bucher_2010,BinnedEST_2016}. In comparison to other methods, such as the KSW estimator~\cite{Komatsu_2005}, this has the advantage of being able to explore the full structure of the calculated bispectra before fitting to any specific template. In this section we outline our implementation of the estimator which follows closely that described in \citet{Coulton_2019}. 

The binned estimator works by creating a set of filtered maps according to some binning scheme: $\{\ell_{\mathrm{min}},\ell_2,...,\ell_{\mathrm{max}}\}$. The $i$th filtered map, $M_{i}(\hat{\bm{n}})$, is then
\begin{align}
M_i(\hat{\bm{n}})=\sum_{l\in\Delta_i}\sum_m Y_{lm}(\hat{\bm{n}})a_{lm},
\end{align}
where $\Delta_i$ is the $i$th bin: $\ell_i\leq l < \ell_{i+1}$. The binned (reduced) bispectrum estimator between three maps, $X$, $Y$ and $Z$, is then given by\footnote{Here, $X$, $Y$ and $Z$ might refer to maps of different foreground components.}
\begin{align}
\hat{b}^{XYZ}_{i,j,k}=\frac{1}{N_{i,j,k}}\int M_i^X(\hat{\bm{n}})M_j^Y(\hat{\bm{n}})M_k^Z(\hat{\bm{n}})\, d^2\hat{\bm{n}} .
\end{align}
Here, $N_{i,j,k}$ is the estimator normalisation, which counts the number of $l$ triplets within the bin triplet ($i,j,k$) satisfying the triangle inequality. It can be expressed in terms of the Wigner-$3j$ symbol as
\begin{multline}
N_{i,j,k}=\sum_{
\ell_1\in\Delta_i,
\ell_2\in\Delta_j,
\ell_3\in\Delta_k}\dfrac{(2\ell_1+1)(2\ell_2+1)(2\ell_3+1)}{4\pi} \\
\times \begin{pmatrix} \ell_1&\ell_2&\ell_3\\
0&0&0
\end{pmatrix}^2 .
\end{multline}
The binned bispectrum estimator provides an unbiased estimate of the reduced bispectrum for the fields $X$, $Y$ and $Z$, defined similarly to equation~\eqref{eq:reducedbispectrum} by
\begin{align}
\langle a^X_{\ell_1 m_1} a^Y_{\ell_2 m_2} a^Z_{\ell_3 m_3} \rangle =b^{XYZ}_{\ell_1\ell_2\ell_3}\mathcal{G}^{m_1m_2m_3}_{\ell_1\ell_2\ell_3},
\end{align}
in the limit that the reduced bispectrum does not vary appreciably across the bin widths.
The estimator only counts modes that satisfy the triangle condition and have $\ell_1+\ell_2+\ell_3=\text{even}$. The estimator normalisation thus effectively counts the number of triangles also satisfying the parity selection rule in the bin triplet. This parity selection rule is a consequence of an underlying assumption that the (joint) probability distribution function of the fields is invariant under spatial inversion.

Note that the bispectrum is invariant under simultaneous permutations of the map labels $X$, $Y$ and $Z$ and the bin labels (or multipoles for the unbinned case) $i$, $j$, $k$. We identify the distinct elements through the restriction $i\leq j \leq k$, in which case it is generally necessary to distinguish between permutations of the fields.

If we assume that the fields are weakly non-Gaussian, we can use Wick's theorem to get the covariance of the estimator: 
\begin{multline}
\text{Cov}\left(\hat{b}^{XYZ}_{i,j,k},\hat{b}^{X'Y'Z'}_{i',j',k'}\right) = \frac{1}{(N_{i,j,k})^2} \\
\times \sum_{
\ell_1\in\Delta_i,
\ell_2\in\Delta_j,
\ell_3\in\Delta_k}
\frac{(2\ell_1+1)(2\ell_2+1)(2\ell_3+1)}{4\pi} {\begin{pmatrix} 
\ell_1&\ell_2&\ell_3\\
0&0&0
\end{pmatrix}
}^2 \\
\times \left[C^{X,X'}_{\ell_1}C^{Y,Y'}_{\ell_2}C^{Z,Z'}_{\ell_3} \delta_{i,i'} \delta_{j,j'} \delta_{k,k'} + \text{5 perms}\right],
\label{eq:bispectrumVar}
\end{multline}
where the permutations are over the set with elements $\{ X' i', Y' j', Z' k'\}$. Of these permutations, six, two or one are non-zero depending on whether three, two or none of the multipole bins are equal, respectively. Note that bispectrum estimators with distinct bin labels $i\leq j \leq k$ are treated as uncorrelated. $C^{X,X'}_{\ell}$ is the (cross-)power spectrum between $X$ and $X'$ defined as 
\begin{equation}
 \langle a^{X}_{\ell m}  \left(a^{X'}_{\ell' m'}\right)^\ast \rangle =     C^{X,X'}_{\ell}  \delta_{m,m'}\delta_{\ell,\ell'}.
\end{equation}
The results of the binned bispectrum estimator obviously depend on the choice of binning. In this paper we use the following binning scheme with $\ell_{\mathrm{max}}=2960$:
\begin{center}
[2,   25,   52,  128,  212,  272,  332,  392,  452,  524,  644, 764,  908, 1028, 1172, 1292, 1412, 1556, 1676, 1820, 1940, 2110, 2210, 2360, 2460, 2610, 2710, 2860, 2960].
\end{center}
This binning scheme is somewhat arbitrary but has been chosen with some important features in mind. First, there are more bins on large scales to capture the stronger scale dependence of squeezed configurations. This is also important as the ISW effect drops off rapidly with $l$. On smaller scales the signals are smoother and thus larger bins can be used. Fortunately in either case, the optimally of the estimator does not depend strongly on the choice of binning and so we can be confident in the accuracy of our results (see \citealt{BinnedEST_2016} and \citealt{Coulton_2020}). 

\subsection{Template fitting}\label{sec:templateFitting}
Estimating $\fnl$ from the binned bispectrum is relatively straightforward \citep{BinnedEST_2016}. The first necessary step is to bin the theoretical template bispectrum (as discussed in Section \ref{sec:backgroundContext}) according to the same binning scheme:
\begin{multline}
b^{\mathrm{th}}_{ijk}=\frac{1}{N_{i,j,k}}\sum_{\ell_1\in\Delta_i}\sum_{\ell_2\in\Delta_j}\sum_{\ell_3\in\Delta_k}b^{\mathrm{th}}_{\ell_1\ell_2\ell_3} \\
\times \frac{(2\ell_1+1)(2\ell_2+1)(2\ell_3+1)}{4\pi} {\begin{pmatrix} 
\ell_1&\ell_2&\ell_3\\
0&0&0
\end{pmatrix}
}^2 .
\end{multline}
The estimator for $\fnl$ from the bispectrum of a single map is then 
\begin{align} \label{eq:fnl_estimator}
\hat{f}_{\mathrm{NL}}&=\dfrac{{b^{\mathrm{th}} \circledast \hat{b}}}{{b^{\mathrm{th}} \circledast b^{\mathrm{th}}}},
\end{align}
where,
\begin{align} 
{b^A \circledast b^B}&=\sum_{i\leq j \leq k}b^{A}_{ijk}(V^{-1})_{i,j,k}b^{B}_{ijk},
\end{align}
which forms an inner product. Here, $V_{i,j,k} = \text{Var}(\hat{b}_{i,j,k})$ and including its inverse in the inner product
provides a minimum-variance estimate of $\fnl$~\citep{Babich_2005,Creminelli2006}. 

If the bispectrum estimate, $\hat{b}_{i,j,k}$, is from maps contaminated with multiple extragalactic foregrounds, it will consist of a sum of contributions with expectation values $b^{XYZ}_{i,j,k}$, where $X$, $Y$ and $Z$ may be lensed CMB or foregrounds. For each choice of components, all distinct permutations of the underlying fields will appear. If not all three fields are CMB, their associated bispectrum will contribute to a bias,
$\Delta\fnl$, on a measurement of primordial non-Gaussianity with
\begin{align} \label{eq:est_bias}
    {\Delta\fnl^{XYZ} = \dfrac{{b^{\mathrm{th}} \circledast b^{XYZ} }}{{b^{\mathrm{th}} \circledast b^{\mathrm{th}}}}}.
\end{align}
The size of this bias will depend on how similar the bispectra are to the primordial templates and we will assess the significance of the bias by comparing it to the variance of the primordial template amplitude
\begin{align} \label{eq:var_fnl_PNG}
    \sigma^2[\hat{f}_{\mathrm{NL}}] = \dfrac{1}{{b^{\mathrm{th}} \circledast b^{\mathrm{th}}}}. 
\end{align}

In this work, we compute biases from all permutations of extragalactic foreground components using equation~\eqref{eq:est_bias}. Since we estimate the bias from a single realisation of the lensed CMB and foregrounds, there is a variance associated with each estimated bias. We calculate this in the Gaussian limit using
\begin{align}\label{eq:bias_variance}
  &  \mathrm{Var}[\Delta\fnl^{XYZ}] = \nonumber \\ & \dfrac{1}{{(b^{\mathrm{th}} \circledast b^{\mathrm{th}})^2}} \sum_{i\leq j \leq k}b^{\mathrm{th}}_{ijk}(V^{-1})_{i,j,k} \tilde{V}^{XYZ}_{i,j,k} (V^{-1})_{i,j,k}b^{\mathrm{th}}_{ijk},
\end{align}
where $\tilde{V}^{XYZ}$ is the Gaussian variance of the bispectrum $\hat{b}^{XYZ}_{i,j,k}$ (including all distinct permutations of the sky components $X$, $Y$ and $Z$), which can be calculated using equation~\eqref{eq:bispectrumVar}.
Note that $\tilde{V}^{XYZ}_{i,j,k}$ is much smaller than the variance from all the components, $V_{i,j,k}$, and thus we have a lower variance on the inferred biases than the estimator variance given in equation~\eqref{eq:var_fnl_PNG}. However by using the Gaussian variance we are underestimating the variance of the bias as, for many of these sources, there will be significant non-Gaussian contributions to the variance (e.g., from the connected four- and six-point functions).

We will consider three different types of measurement, each with different estimator weightings, i.e., different $V_{i,j,k}$ in the inner products in equation~\eqref{eq:fnl_estimator}: first we consider cosmic-variance-limited bispectrum measurements, where the CMB power spectra used to compute $V_{i,j,k}$,  equation~\eqref{eq:bispectrumVar}, consist only of the primary lensed CMB. This is only used in Section~\ref{sec:ISW+Lensing_biases} to compare to the results in \citet{Hill_2018}. Second, in Section \ref{sec:total_bias}, we consider measurements at a single frequency; here the power spectra in equation~\eqref{eq:bispectrumVar} will have contributions from the CMB and extragalactic foregrounds at that frequency and instrumental effects including noise and beam.  Finally, in Section \ref{sec:CleanedResults}, we consider component-separated measurements, where the power spectra are obtained from a weighted sum of measurements at a range of frequencies.

\section{Secondary Anisotropies}\label{sec:secondaryAnisotropies}

The observed intensity of the microwave sky is composed of the primary CMB contaminated with Galactic emission and several extragalactic signals called secondary anisotropies. We investigate how the non-Gaussianity of these foregrounds can give rise to non-primordial bispectra and bias our measurements of PNG. In this section, we briefly review the physics behind these effects and examine why we expect them to bias to PNG measurements.

\subsection{Gravitational lensing} 
As the CMB photons propagate from the surface of last scattering to observation, their paths are deflected by the inhomogenous distribution of matter in the universe. This remaps our view of the CMB as~\citep{Seljak_1996,LEWIS_2006}
\begin{equation}\label{eq:lensing}
\widetilde{T}(\hat{\boldsymbol{n}})=T(\hat{\boldsymbol{n}}+\boldsymbol{\nabla}\phi),
\end{equation} 
where $\widetilde{T}$ is the lensed temperature and $\boldsymbol{\nabla}\phi$ is the gradient of the lensing potential, given in a flat universe by
\begin{equation} \label{eq:lensingPot}
    \phi(\hat{\boldsymbol{n}}) = -2 \int\limits_0^{\chi_*}\mathrm{d}\chi \frac{\chi_*-\chi}{\chi_*\chi} \Phi(\chi\hat{\boldsymbol{n}},\chi),
\end{equation}
where $\Phi$ is the Newtonian gravitational potential, $\chi$ is the conformal distance and $\chi_*$ is the conformal distance to the surface of last scattering. Lensing is often quantified in terms of the convergence, $\kappa$, which is half the Laplacian of the lensing potential, $\kappa=-\frac{1}{2}\nabla^2\phi$. 
The convergence (and lensing potential) entails information about the integrated line-of-sight mass density field of the LSS. Gravitational lensing does not alter the black body spectrum of the CMB and thus cannot be removed via multi-frequency analysis.

\subsection{The integrated Sachs--Wolfe (ISW) effect}
The integrated Sachs--Wolfe (ISW) effect occurs when photons become redshifted as they propagate through evolving gravitational potentials \citep{Sachs_1967}. In this work we are primarily concerned with the linear, late-time ISW effect. This occurs as photons gain, or lose, energy as the large-scale linear potentials and voids decay due to the late-time accelerated
expansion of the universe. This effect generates CMB anisotropies that trace the potential fields as
\begin{align}\label{eq:ISW}
\frac{\Delta T^{\mathrm{ISW}}}{T_\mathrm{CMB}}=2\int_{\eta_*}^{\eta_0}d\eta\frac{\partial\Phi}{\partial\eta},
\end{align}
where $\Phi$ is the Newtonian potentials. 
The ISW effect generates temperature anisotropies with the same frequency spectrum as the primary fluctuations and
and thus cannot be separated from them with multi-frequency cleaning. The late-time ISW effect is primarily important at low redshifts, $z\lesssim 2$, and will be correlated with the other secondary anisotropies as they trace the matter over-density (and hence the potential).

Note that the ISW effect also occurs near recombination, as the universe completes the transition from radiation domination to matter domination. Non-Gaussianity related to this is known as the intrinsic bispectrum \citep[see, e.g.,][]{Bartolo_2006,Lewis_2012,Pettinari_2013,Pettinari_2016,Coulton_2021} . The ISW effect also occurs on smaller scales, where potentials grow due to non-linear evolution. This effect is known as the non-linear ISW, or Rees--Sciama effect \citep{ISW_Rees-Sciama_1968}. We do not consider this effect as it does not generate significant biases of PNG searches due to its small magnitude and scale dependence.

\subsection{The Sunyaev--Zel'dovich effect}
CMB photons that are scattered by free electrons in galaxy clusters generate new anisotropies. This effect is known as Sunyaev--Zel'dovich effect and is further split into two parts: kinetic (kSZ) and thermal (tSZ). 
\subsubsection{The thermal Sunyaev--Zel'dovich effect}
Hot electrons in the ionized intracluster medium of galaxy clusters can transfer energy to CMB photons by inverse Compton scattering. The strength of the tSZ effect is characterized by the Compton-$y$ parameter, $y$, which is a line-of-sight integral of the electron pressure, of the photons according to \citep{Sunyaev_1972,Sunyaev_1980}:
\begin{align} \label{eq:tSZ}
\frac{\Delta T^{\mathrm{tSZ}}}{T_{\mathrm{CMB}}} &= f(x)y=\left[x\coth\left(\dfrac{x}{2}\right)-4\right]\int an_e\sigma_{\text{T}}\frac{k_\text{B}T_e}{m_ec^2}\, d\chi,
\end{align}
where $a$ is the scale factor, $n_e$ is the electron number density, $\sigma_{\text{T}}$ is the Thompson scattering cross section, $k_{\text{B}}$ is the Boltzmann constant, $T_e$ is the electron temperature $m_ec^2$ is the electron rest mass energy, and $x=h\nu/(k_{\text{B}}T_{\text{CMB}})$ is the dimensionless frequency. The tSZ effect is a frequency-dependent shift of the photons, and thus changes the blackbody spectrum of the CMB. Importantly, $f(x)=0$ for $\nu\approx 217\,\text{GHz}$ and so the tSZ signal vanishes when viewing the sky at $217\,\text{GHz}$ (one of the HFI channels used by \textit{Planck}). This is important for the multi-frequency analysis.

\subsubsection{The kinetic Sunyaev--Zel'dovich effect}\label{sec:ksz}
The kSZ effect arises from the scattering of photons from electrons in bulk motion due to peculiar velocities with
respect to the reference frame of the CMB. For motion with line-of-sight velocity $v_{||}$ the temperature change is \citep{Sunyaev_1972,Sunyaev_1980}
\begin{align}\label{eq:kSZ}
\frac{\Delta T^{\mathrm{kSZ}}}{T_{\mathrm{CMB}}}=\sigma_{\text{T}}\int  d\chi e^{-\tau}n_e(\chi)\frac{v_{||}}{c}a(\chi),
\end{align}
where $\tau$ is the Thomson scattering optical depth
\begin{align}
\tau=\int \mathrm{d}\chi\, a(\chi) n_e(\chi) \sigma_{\text{T}} .
\end{align}
The kSZ effect does not alter the blackbody spectrum of the CMB and so cannot be mitigated via multi-frequency analysis. An important property of this effect is that all moments containing an odd number of the field vanish. This is because, from homogeneity, we expect to see an equal number of structures moving towards us as we would see moving away from us, when averaged over the whole sky. This means that we only expect non-zero bispectra involving exactly two (or zero) kSZ legs. 

\subsection{Radio galaxies}
In contrast to the previous effects, which act directly on the CMB, radio galaxies present a contaminating emission on top of the CMB. Synchrotron radiation from the jets and radio lobes of active galactic nuclei (AGN) leads to bright point-source anisotropies in CMB maps, known as radio galaxies \citep[see, e.g.,][for a review]{deZotti2010}. Radio galaxies are typically divided into two classes: steep-spectrum sources, with emission primarily from radio lobes, and flat-spectrum sources, with emission primarily from the AGN jets \citep{Fanaroff_1974}. Radio galaxies are the dominant extragalactic foreground at low frequencies, $\nu \lesssim 90\,\text{GHz}$, and they become less important at higher frequencies. 

\subsection{The cosmic infrared background (CIB)} 
The final effect we consider is emission from high-redshift, dusty star-forming galaxies. Like radio galaxies, this is a contaminating emission on top of the CMB. This emission arises from the absorption of UV radiation from young stars by dust in the host galaxy. The dust then re-radiates at longer wavelengths \citep{Puget_1996,Guiderdoni_1997,Puget_1999,Gispert_2000,Lagache_2006}. In the source rest-frame the emission follows a modified blackbody distribution \citep[see, e.g.,][]{Marsden_2011,Negrello_2007} and results in anisotropies with a unique spectral dependence that peaks around $2\,\text{THz}$. Thus, the CIB is increasingly important at high frequencies.

\subsection{Biasing of primordial non-Gaussianity} \label{sec:bias_intuition} 

Before discussing our measurements on simulations, it is instructive to consider what kinds of bias we expect. To do this we consider biases on local-type non-Gaussianity measurements.  
The local bispectrum arises from a quadratic modification to a purely Gaussian curvature perturbation, $\zeta_{G}$:
\begin{align}
\zeta=\zeta_{G}+\frac{3}{5}\fnlloc\left(\zeta_{G}^2-\braket{\zeta_G^2}\right) .
\end{align}
This has the particular effect of modulating the small-scale power by the large-scale modes of $\zeta_{G}$. Specifically, we have the following approximation for the local small-scale power:
\begin{align}
P_{\zeta,\text{small}}(k)&\approx P_{\zeta_G,\text{small}}(k)\left(1+\frac{6}{5}\fnlloc\zeta_{G,\mathrm{large}}\right) .
\end{align}
Using the Sachs--Wolfe relation between large-scale temperature anisotropies (at the spatial origin) and curvature perturbations,
\begin{equation}
\frac{\Delta T(\hat{\boldsymbol{n}})}{T_{\text{CMB}}} = - \frac{1}{5}\zeta(\chi_* \hat{\boldsymbol{n}}),
\end{equation}
we can estimate the impact on the locally measured small-scale CMB power spectrum as
\begin{align}
C_{\ell}(\zeta_{\mathrm{small}}) &\approx C_{\ell}(\zeta_{G,{\mathrm{small}}}) \left(1-6\fnlloc\left(\frac{\Delta T}{T_{\text{CMB}}}\right)_{G,\mathrm{large}}\right) .
\end{align}
This allows us to interpret physically the sign of $\fnlloc$ as follows. 
\begin{itemize}
\item Positive $\fnlloc$: less small-scale power in regions of large-scale $\Delta T>0$ and more small-scale power in regions of large-scale $\Delta T<0$.
\item Negative $\fnlloc$: more small-scale power in regions of large-scale $\Delta T>0$ and less small-scale power in regions of large-scale $\Delta T<0$.
\end{itemize}

Let us now consider the biases that arise when forming the bispectrum of the total sky signal, which we write as
\begin{equation}
T_{\text{tot}} = T + \Delta T + \sum_i X_i ,
\end{equation}
where $T$ is the (unlensed) primary CMB anisotropy, $\Delta T$ is the change induced by gravitational lensing, i.e.,
\begin{equation}
\Delta T = \tilde{T}-T = \boldsymbol{\nabla} \phi \cdot \boldsymbol{\nabla} T + \cdots ,
\label{eq:lensing_taylor_exp}
\end{equation}
and the $X_i$ are the secondary anisotropies. Forming the bispectrum for the case of Gaussian primary CMB anisotropies, we have (symbolically)
\begin{multline}
\langle T_{\text{tot}} T_{\text{tot}} T_{\text{tot}} \rangle = \left(\langle T \Delta T \sum_i X_i \rangle + \text{5 perms}\right) \\ + \left(\langle \Delta T \Delta T \sum_i X_i \rangle + \text{2 perms}\right) + \langle \sum_i X_i \sum_j X_j \sum_k X_k \rangle \, .    
\label{eq:totbispecexpand}
\end{multline}
The first two terms arise from two of the legs of the bispectrum being the lensed CMB. These terms will generally be non-zero for secondary anisotropies $X_i$ that are correlated with CMB lensing. The most well-known example is the ISW-lensing bispectrum~\citep{Goldberg_1999,Smith_2011,Lewis_2011}.

The impact of lensing by a large-scale over-density is to magnify the small-scale CMB anisotropies; this causes a reduction in small-scale power, as the CMB power spectrum decreases sharply at small scales. Thus for fields, $X_i$, that are positively correlated with the large-scale lensing convergence, we have the same effect as local-type non-Gaussianity with $\fnlloc > 0$: an excess large-scale signal reduces locally the small-scale power.\footnote{This effect is captured by the first term on the right of equation~\eqref{eq:totbispecexpand}.} As an example, the ISW effect towards a large-scale over-density is positive and so is positively correlated with the lensing convergence. The ISW-lensing bispectrum therefore gives rise to a positive bias to $\fnlloc$.

The third term on the right of equation~\eqref{eq:totbispecexpand} arises from the bispectra of the secondary anisotropies. As well as the intrinsic bispectra of the individual components, there can be mixed bispectra. For example, a large-scale over-density produces a positive ISW signal and a local excess of small-scale shot-noise power in the over-density of discrete objects (e.g., galaxy clusters). It follows that the ISW-tSZ-tSZ bispectrum leads to a negative bias to $\fnlloc$.

\section{Simulations} \label{sec:simulations}
In this work we consider two sets of simulations: the Websky simulations \citep{Stein_2020} and the Sehgal et al.~simulations \citep{sehgal_2010}. These two simulations use different, and complementary, methods to simulate maps of CMB secondary anisotropies and by comparing the two we can gauge the theoretical uncertainty in our predictions. In this section we briefly review the details of the two simulations, referring the readers to \citet{Stein_2018,Stein_2020} and \citet{sehgal_2010} for further details.

Note that for both sets of simulations the frequency-dependent components are evaluated at a single frequency rather than integrated with respect to an instrumental bandpass.

\subsection{Websky simulations} \label{sec:simulations_websky}
The Websky simulations comprise one realization of the integrated Sachs--Wolfe effect, thermal and kinetic Sunyaev--Zel'dovich effects, the cosmic infrared background and gravitationally lensed CMB.  Importantly, these components are non-Gaussian and correlated, which is vital for an accurate characterization of any biases on PNG.

The Websky maps are produced from a full-sky simulation of the matter distribution over the range $0<z<4.5$ with a mass resolution of $\sim 10^{12}\,\mathrm{M}_\odot h^{-1}$, and is composed of two components: the halo and the field components. The halo component consists of a catalog of halos and their associated masses. This catalog is computed using the peak-patch algorithm \citep{Bond_1996}. The field component accounts for the remaining matter that is absent from the halos and is computed using second-order Lagrangian perturbation theory (2LPT). This forms a major part of the lensing and kinetic Sunyaev Zel'dovich effects. The maps of CMB secondary effects are obtained by `painting' these effects onto the halo catalog and field components and then projecting them along a light cone. The cosmology used is the best-fit LCDM model from~\citet{planck2016-l06}.

\subsubsection{Gravitational lensing}
Gravitaionally-lensed maps of the CMB are computed in two stages. First the lensing potential is computed from the matter over-density field using equation~\eqref{eq:lensingPot} and the Poisson equation. The contribution to the matter-over density field from the halos is calculated by assuming that the matter in the halos follows a modified NFW profile \citep{Navarro1996}. To avoid double counting from the field and halo contributions at the locations of the halos, a uniform density sphere of matched mass is subtracted at the location of each halo. The lensing potential is then used to deflect a Gaussian CMB generated from CAMB~\citep{Lewis:1999bs} using equation~\eqref{eq:lensing}. The lensing map is constructed using the Born approximation.

\subsubsection{Integrated Sachs--Wolfe effect}
The ISW field is computed using equation~\eqref{eq:ISW} and the linear-theory Newtonian potentials. 
The derivatives of the potentials are given by 
\begin{equation}
    2\frac{\partial\Phi(\boldsymbol{x},\eta)}{\partial\eta} = -3 \Omega_{m,0}\left( \frac{H_0}{k}\right)^2     \frac{\partial (D/a)}{\partial\eta}\delta^L(\boldsymbol{x},\eta=0),
\end{equation}
where $D$ is the linear growth function for the matter overdensity, which is proportional to $a$ in matter domination, and $\delta^L(\boldsymbol{x},\eta=0)$ is the linear initial-condition field. Note that this method means that there is no non-linear contributions to the ISW field. These are expected to be small \citep{Cooray_2002} and so are unlikely to impact our results.

\subsubsection{Thermal Sunyaev--Zel'dovich effect}

The tSZ effect is simulated through modelling the electron pressure distribution throughout the universe and then computing the tSZ effect via equation~\eqref{eq:tSZ}. The pressure is modelled by assigning spherical pressure profiles to the halos. The pressure profiles are those reported in \citep{Battaglia_2012} and are obtained from an analytical fit to the distribution of gas in hydrodynamical simulations of galaxy clusters. The non-Gaussian nature of the tSZ effect arises from the non-Gaussian distribution of galaxy clusters, with a part that is Poisson-like, from the discrete nature of the halos, and a clustering part as the halos trace the matter overdensity field. Note there is no contribution from the field component, which is an accurate approximation as the tSZ effect gets the majority of its contribution from massive halos \citep{Komatsu_2002b}.

\subsubsection{Kinetic Sunyaev--Zel'dovich effect}

The kSZ effect gains contributions from both the field and halo components. The halo contribution is implemented by first assigning an electron density distribution to each halo; the profiles used are the analytical fits to hydrodynamic simulations in \citet{Battaglia_2016}. The halo electron density is then combined with the dark matter halo's velocity and projected along the line-of-sight as in equation~\eqref{eq:kSZ}. A similar procedure is done to compute the field contribution: the electron density is assumed to be a unbiased tracer of the underlying matter density, with a multiplicative correction to account for the number of ionized electrons. Thus, the field contribution is straightforwardly converted to the electron overdensity. This is combined with a velocity field obtained from 2LPT and then projected along the light cone.

\subsubsection{Cosmic infrared background}

The CIB is modelled in the Websky simulations by assigning galaxies to the dark matter halos based on a halo occupation distribution \citep{Zheng2005,Zheng_2009}. For each halo, the expected number of galaxies per halo is calculated according to the halo's mass and redshift. The actual number of galaxies assigned to that halo is drawn from a Poisson distribution with the appropriate mean.
These galaxies are then randomly spatially distributed within the halo with a spectrum and luminosity determined by a model based on measurements by the \textit{Herschel Space Observatory} \citep{Viero_2013}, with an amplitude updated to match the measurements by the \textit{Planck} satellite \citep{planck2013-pip56}. Note that the CIB is formed of sources with similar rest-frame emission spectra but sourced at different redshifts. This results in anisotropies that are not perfectly correlated across different frequencies, an effect that is included in the Websky simulations. As discussed in Appendix~\ref{app:PSmasking}, we found it necessary to mask the brightest sources and as such we mask all sources with a flux density greater than 15\,mJy at 217\,GHz.

\subsubsection{Radio galaxies}
At the time of this work, no radio galaxy maps were available for the Websky simulations. Very recently \citet{Li_2021} have produced a set of radio galaxy maps for this simulation. We leave an examination of this to future work.

\subsection{Sehgal et al.~simulations} \label{sec:simulations_sehgal}

The Sehgal et al.~simulations \citep{sehgal_2010} provide one realization of the thermal and kinetic Sunyaev--Zel'dovich effects, the lensed CMB, CIB and radio galaxies. We use the updated Sehgal et al.~maps presented in \citet{SO_2019}. As \citet{Hill_2018} found that there are interesting potential biases from correlations with the ISW effect we have constructed an approximate version of the ISW field for the Sehgal et al.~simulations. The details of this construction are provided in Appendix~\ref{app:ISW-sehgal}.

Similar to the Websky simulations, the CMB secondary anisotropies are `painted' onto a simulation of the dark matter. In this case the dark matter is simulated via a tree-particle-mesh code \citep{Bode_2000,Bode_2003} with a periodic box of side 1\,Gpc$^{-1}$, $1024^3$ particles and a mass resolution of $ 6.82\times10^{10}\,\mathrm{M}_\odot h^{-1}$, and run at the WMAP-5 cosmology \citep{komatsu_2009}. Halos are identified with the friends-of-friends algorithm \citep{Huchra_1982,Nolthenius_1987}. One octant of the light cone is obtained from slices of the periodically tiled box. For $z<3$, the particles are used whereas for $10>z>3$ a pixellated version of the particle data is used. This octant is then replicated to obtain a full-sky map. As described below, some of the painting procedures are significantly different to the Websky case, which enables us to assess the theoretical uncertainty in our predictions. 

\subsubsection{Gravitational lensing}

The construction of the lensing map is generally similar to the Websky case with the exception that the mass over-density field is constructed from the particle data, rather than the two components of the Websky simulation. The full algorithm is described in \citet{Das_2008} and, like Websky, uses the Born approximation.

\subsubsection{Thermal Sunyaev--Zel'dovich effect}

The tSZ effect is computed by modelling the pressure field in combination with equation~\eqref{eq:tSZ}. The pressure field is composed of three separate components: a contribution from massive halos; from low-mass halos and the intergalactic medium (IGM); and the contribution from high redshift ($z>3$). The halo component is the dominant contribution to the tSZ effect and is modelled by a method based on the assumption of hydrostatic equilibrium, with parameters tuned to match X-ray observations \citep{Bode_2009}. Unlike the Websky model, this captures the effects of substructure and triaxiality. The low-mass halo and IGM contributions are computed by using the virial theorem to estimate the pressure as there is insufficient resolution to apply the hydrostatic equilibrium model. The high-redshift part is computed from the pixellated density, combined with a fixed temperature field.  

As described in \citet{SO_2019}, we rescale the tSZ map by a factor of 0.75 to match observations from \textit{Planck} \citep{planck2014-a28}, ACT \citep{Dunkley_2013} and SPT \citep{George_2015} better.

\subsubsection{Kinetic Sunyaev--Zel'dovich effect}

The kSZ effect is modelled by combining the electron momentum field with equation~\eqref{eq:kSZ}. The physical model is similar to the tSZ: there is a contribution from massive halos that is modelled by the hydrostatic equilibrium model, to compute the electron density, combined with a mass-weighted halo average velocity. 
The contributions from low-mass halos and the IGM are computed using the virial theorem and the high-redshift component is computed using the pixellated density and momentum fields. Usually the kSZ effect has large cancellations along the line of sight. These are missing due to the repeated tiling of the box so radial filtering is used to remove this extra power. This filtering means that the kSZ is only accurate for multipoles $\ell\gtrsim 1000$ and that the Sehgal et al.~kSZ bispectra are likely underestimated. However, as we find biases involving the kSZ to be completely negligible, even in Websky, this is not a significant limitation of our analysis. 

\subsubsection{Cosmic infrared background}

Like the Websky simulations, the CIB is implemented by a halo occupation distribution. In this case there are two populations of galaxies, a faint and a bright population. Given the expected number of galaxies for each halo, a realization is drawn from a Poisson distribution. The luminosity of the galaxies depends on the host halo mass and redshift, based on a model from \citet{Righi_2008} with updated parameters to fit a range of number count and clustering measurements, with an additional stochastic component in the luminosity and dust temperature. To avoid losing the contribution to the CIB from halos below the minimum resolved halo mass, $6.82\times 10^{12}\,\mathrm{M}_\odot h^{-1}$, an additional component is added to the simulations. An additional number of halos is drawn from the \citet{Jenkins_2001} mass function. These are then Gaussianly distributed around the lowest-mass resolved halos. These halos are then populated with CIB galaxies as above.

In this work we use the updated version of the CIB maps; see \citet{SO_2019} for more details. These utilize information from small-scale measurements of the CIB by \citet{Dunkley_2013} to improve the model's accuracy and are interpolated for the frequencies considered in this work. Note there are no CIB maps for the 545 GHz \textit{Planck} channel, which slightly impacts our component separation, as discussed in Section \ref{sec:CleanedResults}. As in the Websky case, we also mask bright CIB sources with flux density greater than 15\,mJy at 217\,GHz.

\subsubsection{Radio galaxies}

The radio galaxies are also simulated using a halo occupation distribution. Two populations of radio galaxies are randomly drawn from a Poisson distribution and with luminosities randomly drawn from a model based on modelling by \citet{Wilman_2008} and measurements from \citet{Lin_2009}. The two populations approximately correspond to the two classifications
of \citet{Fanaroff_1974} and the predicted properties match number counts at a range of frequencies and redshifts.

We use the modified maps that have had a 7\,mJy flux-density cut applied to them and interpolated for the frequencies considered here, as described in \citet{SO_2019}. For mock observations of a \textit{Planck}-like experiment , these cuts remove too many of the sources given the experiments resolution, however the brightest objects are very rare so their clustering signals are dominated by shot-noise.
Thus we expect the modified maps still to provide a reasonable approximation of the biases that arise from radio galaxies.

\section{Primordial non-Gaussianity biases from single-frequency maps}\label{sec:resultsFreqChan}

We first consider biases to measurements of local, equilateral and orthogonal non-Gaussianties for observations at a single frequency. We consider both a \textit{Planck}-like experiment up to $\ell_{\text{max}}=1590$ or $\ell_{\text{max}}=1940$ with the frequency channels at 100, 143, 217 and 353\,GHz, as well as a forecast for the upcoming Simons Observatory (SO) telescope up to $\ell_{\text{max}}=2960$ for frequency channels at 90, 153, 225 and 278\,GHz~\citep{SO_2019}. For single-frequency observations, and hence no foreground cleaning, we expect to find that there are biases comparable to the statistical errors for both \textit{Planck} and SO-like experiments. In Section \ref{sec:ISW+Lensing_biases} we isolate the subset of bispectra that are expected to be well modelled analytically and give significant biases to $\fnlloc$. We compare our results to those of \cite{Hill_2018} as a consistency check of our simulation methodology. In Section~\ref{sec:total_bias} we consider the full array of biases arising from different combinations of all foregrounds for \textit{Planck}-like and SO-like experiments.

\subsection{Analytic biases to local-type non-Gaussianity measurements}\label{sec:ISW+Lensing_biases}

\begin{table*}
 \begin{tabular}{c  c c c c  c} 
 \hline
 & \multicolumn{2}{c}{ $\ell_{\text{max}}=1590$ (\textit{Planck}-like)}& & \multicolumn{2}{c}{ $\ell_{\text{max}}=2960$ (SO-like)} \\
  &bias to $ \fnlloc$ & bias to $\fnlloc (\text{Hill})$ &  & bias to $ \fnlloc$  & bias to $\fnlloc (\text{Hill})$\\
 \hline
100\,GHz &&& 100\,GHz && \\ 
 \hline  
ISW-lensing & 9.7 $\pm$ 2.0 & 7.6 & ISW-lensing & 15.1 $\pm$ 1.1 & 14.9  \\  
 CIB-lensing & 0.9 $\pm$ 0.2 & 0.9 & CIB-lensing & 1.6 $\pm$ 0.4 & 2.2  \\  
 tSZ-lensing  & -1.0 $\pm$ 0.4 & -1.2  & tSZ-lensing & -1.9 $\pm$ 0.7 & -3.4 \\  
 \hline
143\,GHz &&& 143\,GHz && \\ 
 \hline  
 ISW-lensing & 9.7 $\pm$ 2.0 & 7.6  & ISW-lensing & 15.1 $\pm$ 1.1 & 14.9   \\  
 CIB-lensing & 1.8 $\pm$ 0.3 & 1.4  & CIB-lensing & 3.3 $\pm$ 0.8 & 3.3    \\  
 tSZ-lensing  & -0.7 $\pm$ 0.3 & -0.8 & tSZ-lensing &-1.3 $\pm$ 0.5 & -2.4 \\  
 \hline  
217\,GHz &&& 217\,GHz && \\ 
 \hline  
ISW-lensing & 9.7 $\pm$ 2.0 & 7.6 & ISW-lensing & 15.1 $\pm$ 1.1 & 14.9  \\  
 CIB-lensing & 5.0 $\pm$ 1.0 & 4.3 & CIB-lensing & 9.8 $\pm$ 2.3 & 10.2 \\  
 tSZ-lensing  & $\approx0$ & $\approx0$ & tSZ-lensing & $\approx 0$ &$ \approx 0$   \\
 \hline
353\,GHz &&& 353\,GHz && \\ 
 \hline  
 ISW-lensing & 9.7 $\pm$ 2.0 & 7.6 & ISW-lensing & 15.1 $\pm$ 1.1 & 14.9  \\  
 CIB-lensing & 32 $\pm$ 6.1 & 38 & CIB-lensing & 65 $\pm$ 15 & 88  \\  
 tSZ-lensing  & 1.6 $\pm$ 0.7 & 1.8 & tSZ-lensing & 3.0 $\pm$ 1.2 & 5.0 \\   
 \hline
 \end{tabular}

\caption{Biases to measurements of local-type non-Gaussianity from correlations between gravitational lensing and the  ISW, CIB and tSZ fields for cosmic-variance limited measurements with maximum multipoles of $\ell_{\text{max}}=1590$ and $\ell_{\text{max}}=2960$. We compare results from the Websky simulations to those computed analytically in \citet{Hill_2018}. The errors on the biases are the Gaussian error as calculated by equation~\eqref{eq:bias_variance}. For reference the experimental errors on how well $\fnlloc$ can be measured for each setup are $\sigma=5.3$ for $\ell_{\text{max}}=1590$ and 
$\sigma=2.6$ for $\ell_{\text{max}}=2960$. Only biases comparable to these errors are a concern for experiments.}
\label{tab:compare_lensing}
\end{table*}

\begin{table*}
 \begin{tabular}{c  c c c c c} 
 \hline
  & \multicolumn{2}{c}{ $\ell_{\text{max}}=1590$ (\textit{Planck}-like)}& & \multicolumn{2}{c}{ $\ell_{\text{max}}=2960$ (SO-like)} \\
  & bias to $\fnlloc$  &  bias to $\fnlloc (\text{Hill})$ &  & bias to $ \fnlloc$ &  bias to $\fnlloc (\text{Hill})$\\ \hline
100\,GHz &&& 100\,GHz && \\ 
 \hline  
 ISW-tSZ-tSZ & -1.5 $\pm$ 0.03 & -4.5 & ISW-tSZ-tSZ & -12.9 $\pm$ 0.2 & -47.1 \\
 ISW-CIB-CIB & $\approx 0$ & $\approx 0 $& ISW-CIB-CIB & -0.16 $\pm$ 0.06 & $\approx 0$ \\
 ISW-kSZ-kSZ & -0.05 $\pm$ 0.01 & -0.1 & ISW-kSZ-kSZ & -0.4 $\pm$ 0.02 & -1.7 \\  
 \hline
143\,GHz &&& 143\,GHz && \\ 
 \hline  
 ISW-tSZ-tSZ & -0.75 $\pm$ 0.01 & -2.1 & ISW-tSZ-tSZ & -6.2 $\pm$ 0.1 & -22.5  \\
 ISW-CIB-CIB & $\approx 0 $ & $\approx 0$& ISW-CIB-CIB &-0.7 $\pm$0.2 & -1.2 \\
 ISW-kSZ-kSZ & -0.05 $\pm 0.01$ & -0.1 & ISW-kSZ-kSZ & -0.4 $\pm$ 0.02 & -1.7 \\ 
 \hline  
217\,GHz &&& 217\,GHz && \\ 
 \hline  
ISW-tSZ-tSZ & $\approx 0$  & $\approx 0$ & ISW-tSZ-tSZ & $\approx 0$ & $\approx 0$\\
 ISW-CIB-CIB & -0.01 $\pm$ 0.15 & -0.4 & ISW-CIB-CIB & -7.0 $\pm$ 2.0 & -10 \\
 ISW-kSZ-kSZ & -0.05 $\pm 0.01$ & -0.1 & ISW-kSZ-kSZ & -0.4 $\pm$ 0.02 & -1.7\\
 \hline
353\,GHz &&& 353\,GHz && \\ 
 \hline  
 ISW-tSZ-tSZ & -3.9 $\pm$ 0.1 & -11 & ISW-tSZ-tSZ & -33 $\pm$ 1 & -110 \\
 ISW-CIB-CIB & -38 $\pm$ 6.13 & -33 & ISW-CIB-CIB & -371 $\pm$ 83 & $\approx -1000$\\
 ISW-kSZ-kSZ & -0.05 $\pm 0.01$ & -0.1 & ISW-kSZ-kSZ & -0.4 $\pm$ 0.02 & -1.7 \\
 ISW-tSZ-CIB & -1.6 $\pm$ 1.1 & $ -0.45$ & ISW-tSZ-CIB & -2.3 $\pm$  1.5  & $\approx -1.5$ \\
 \hline
 \end{tabular}
\caption{Biases to measurements of local-type primordial non-Gaussianity from correlations between the ISW effect and the tSZ, CIB and kSZ effects. The experimental setup is otherwise identical to the results in Table~\ref{tab:compare_lensing}. Note we only report the  ISW-tSZ-CIB results at the single frequency reported in \citet{Hill_2018}}
\label{tab:compare_ISW}
\end{table*}
The bispectra for a subset of the CMB secondary anisotropies have been  analytically modelled or approximated in \citet{Hill_2018}. Specifically, these are bispectra with either two lensed CMB maps, as in the first two terms of equation~\eqref{eq:totbispecexpand} with the first term dominating, and one secondary-anisotropy map, or one ISW map and two other secondary maps. Here we briefly review how these are computed, before comparing the analytical biases to those measured in the Websky simulations. We focus specifically on biases to $\fnlloc$, as was the focus in \citet{Hill_2018}, but note that biases to other templates were also calculated. 

The lensing-related biases are simple to model as they are well approximated by the first term of equation~\eqref{eq:totbispecexpand}, which leads to a reduced bispectrum of 
\begin{align}
    b^\mathrm{lens-X}_{\ell_1\ell_2\ell_3 }& = \Big[\frac{\ell_1(\ell_1+1)-\ell_2(\ell_2+1)+\ell_3(\ell_3+1)}{2}C^{\phi X}_{\ell_1}C^{TT}_{\ell_3} \nonumber \\ & + 5\,\mathrm{permutations} \Big] .
\end{align} This requires knowledge of the cross-correlation between the lensing field and the specific secondary anisotropy ( $C^{\phi X}_{\ell_1}$ -in our case ISW, tSZ, CIB and radio). The ISW-lensing cross-power spectrum was calculated in the Limber approximation \citep{Limber,Extended_Limber}, only considering the late-time, large-scale ISW contributions (as is the same in the simulations).
The tSZ-lensing cross-power spectrum was computed in the halo model, including the two-halo term, as in \cite{Hill_2013,Hill_2014}. The CIB-lensing cross-power spectrum was not modelled and instead was chosen to match the best-fit results from \textit{Planck} at each frequency \citep{planck2013-pip56}. 
Explicit details about these computations can be found in the references provided. Note that we do not consider the kSZ-lensing bispectrum here due to the symmetry, described in Section~\ref{sec:ksz}, whereby bispectra with only one kSZ leg vanish. 

The other set of analytic biases to $\fnlloc$ considered in \cite{Hill_2018} are those involving one ISW map and two CIB, tSZ or kSZ maps, specifically biases from the ISW-CIB-CIB, ISW-tSZ-tSZ, ISW-tSZ-CIB and ISW-kSZ-kSZ bispectra. The physical motivation of why these are expected to produce large $\fnlloc$ biases is that the large-scale ISW field at late times modulates the smaller-scale power of the extragalactic anisotropies: the CIB, tSZ and kSZ effects.  In \citet{Hill_2018}, the results were derived within the halo model and focused on the dominant two-halo term~\citep{Buchalter1999}, with the two extragalactic signals in a single halo. The modelling of these bispectra is not as accurate as the lensing-related biases, where only the cross-power spectra required modelling \citep[see, e.g.,][ for a discussion of the accuracy of the halo model for bispectrum calculations]{Lazanu_2017}. 

Comparisons between the analytic calculations (labelled as `Hill' in the following tables) and measurements from  simulations provide a robustness test for both results. Table~\ref{tab:compare_lensing} shows the biases from the lensing-induced bispectra for two different choices of the maximum multipole retained in the analysis: $\ell_{\textrm{max}}=1590$, which approximates a \textit{Planck}-like set of observations and  $\ell_{\textrm{max}}=2960$ for an SO/CMB-S4 like experiment.
 The errors are calculated using equation~\eqref{eq:bias_variance} and, to compare consistently with \citet{Hill_2018}, we use the cosmic-variance-limited (CV) weighting in the bispectrum estimator (see Section~\ref{sec:templateFitting} for more details). Note that the use of the CV weighting means biases can be larger than for a realistic experiment, as is considered in the next section, because the smallest scales, where the CMB secondaries dominate, become overweighted in the estimator with the CV weighting. We reiterate that this error only includes Gaussian contributions and is expected in many cases to be an underestimate of the true error. In general, we find excellent agreement between our results and those in~\citet{Hill_2018}, providing validation for our simulation-based approach. For conciseness we only show the results from the Websky simulations but note that results from the Sehgal et al.~simulations were also in similar agreement. 

As expected, secondary anisotropies that are positively correlated with lensing produce a positive bias to $\fnlloc$. These include the ISW effect, and the CIB at all frequencies (with the CIB bias increasing with frequency as the signal brightness increases). The tSZ effect is negatively correlated with lensing at frequencies below its null at $217\,\text{GHz}$, as the tSZ towards an over-density appears as a temperature decrement at such frequencies, producing a negative bias. At frequencies above $217\,\text{GHz}$, the tSZ appears as a temperature increment and the bias is positive.

A comparison of the ISW-related local-type bias results to those of \citet{Hill_2018} is displayed in Table~\ref{tab:compare_ISW}. While we observe qualitative agreement, we find noticeable differences in these results. In particular, we find generally lower biases arising form the ISW-kSZ-kSZ and ISW-tSZ-tSZ bispectra in both experiment types. This may be a result of inaccuracies in the halo modelling.  We also find a generally lower bias from the ISW-CIB-CIB bispectrum for the SO/S4 forecast, although this can be explained by the differences in our CIB modelling. Note that all displayed biases are negative, as expected, as the small-scale power of the extragalactic signal is enhanced towards over-densities where the large-scale ISW effect is positive. For the CIB and tSZ components, the dependence of the bias on frequency is stronger than in Table~\ref{tab:compare_lensing} since the extragalactic field now appears in two legs of the bispectrum.

\subsection{Simulation biases}\label{sec:total_bias}

Next we consider all possible bispectrum combinations of foregrounds to consider the complete range of biases induced for a \textit{Planck}-like and SO-like experiment. As the observed CMB temperature map is a linear combination of foregrounds, the resulting overall bias on these experiments will also be a linear combination of each component bias (accounting for permutations). Several combinations here have not yet, to our knowledge, been considered before as they are not expected to give significant biases in the primordial templates, and our results serve as a validation of this assumption. In this section, our estimator weighting includes the contributions from noise and foreground signals to match a realistic experiment. 

For the \textit{Planck}-like analysis we use noise and beams from \citet{planck2016-l01} and consider observations at $100$\,GHz, $143$\,GHz, $217$\,GHz and $353$\,GHz with $\ell_{\textrm{max}}=1940$, which matches the analysis done for the temperature-only data \citep{planck2016-l09}. For the SO-like experiment, we use noise and beams from \citet{SO_2019} and consider measurements at $90$\,GHz, $153$\,GHz, $232$\,GHz and $275$\,GHz with an $\ell_{\textrm{max}}=2960$. For the SO-like experiment we do not include the effect of atmospheric noise at large scales, so our results are representative of combining SO with measurements on large and intermediate scales from \textit{Planck}.

\begin{figure*}
\includegraphics[height=11.0cm,width=15.5cm]{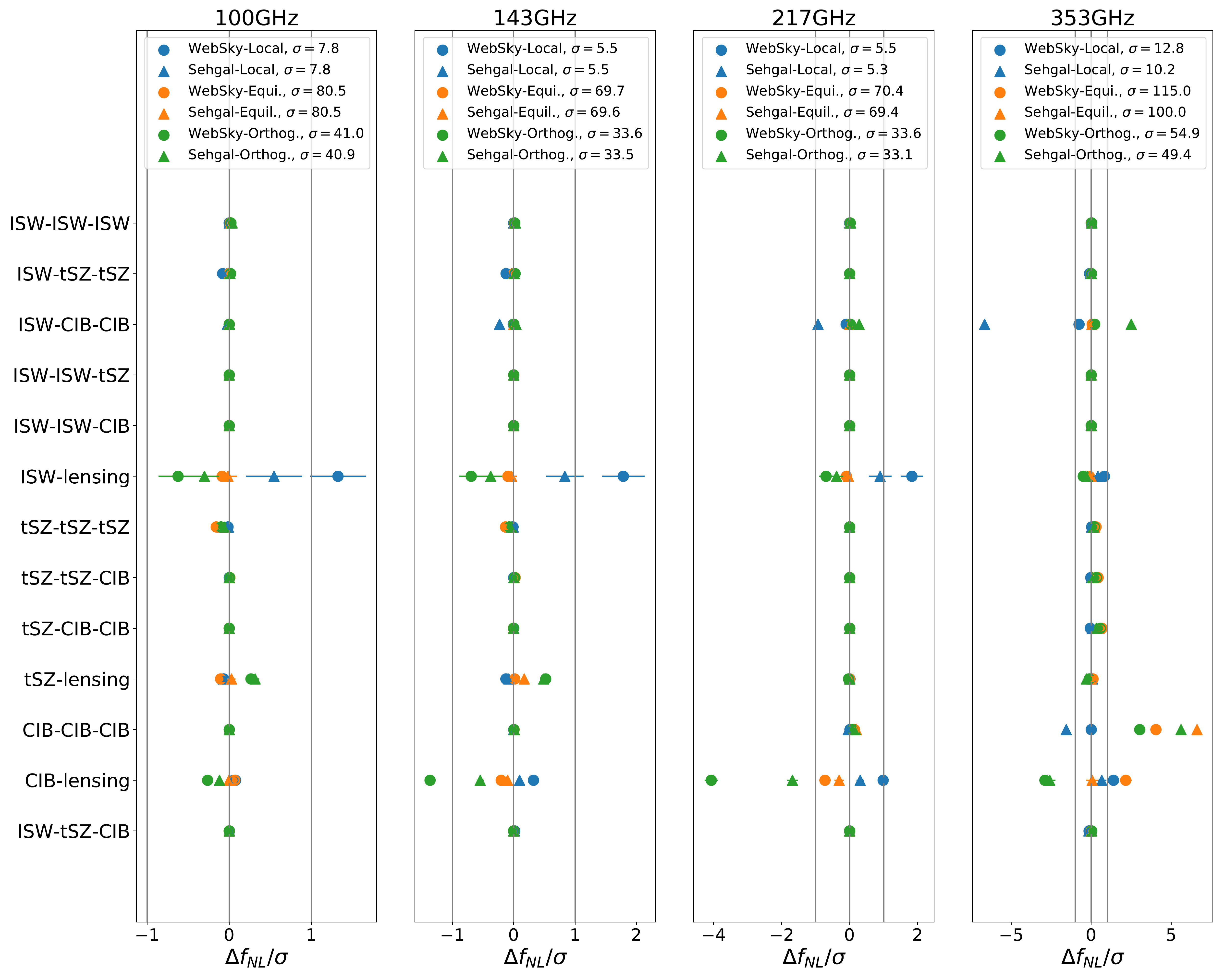}
\includegraphics[height=11.0cm,width=15.5cm]{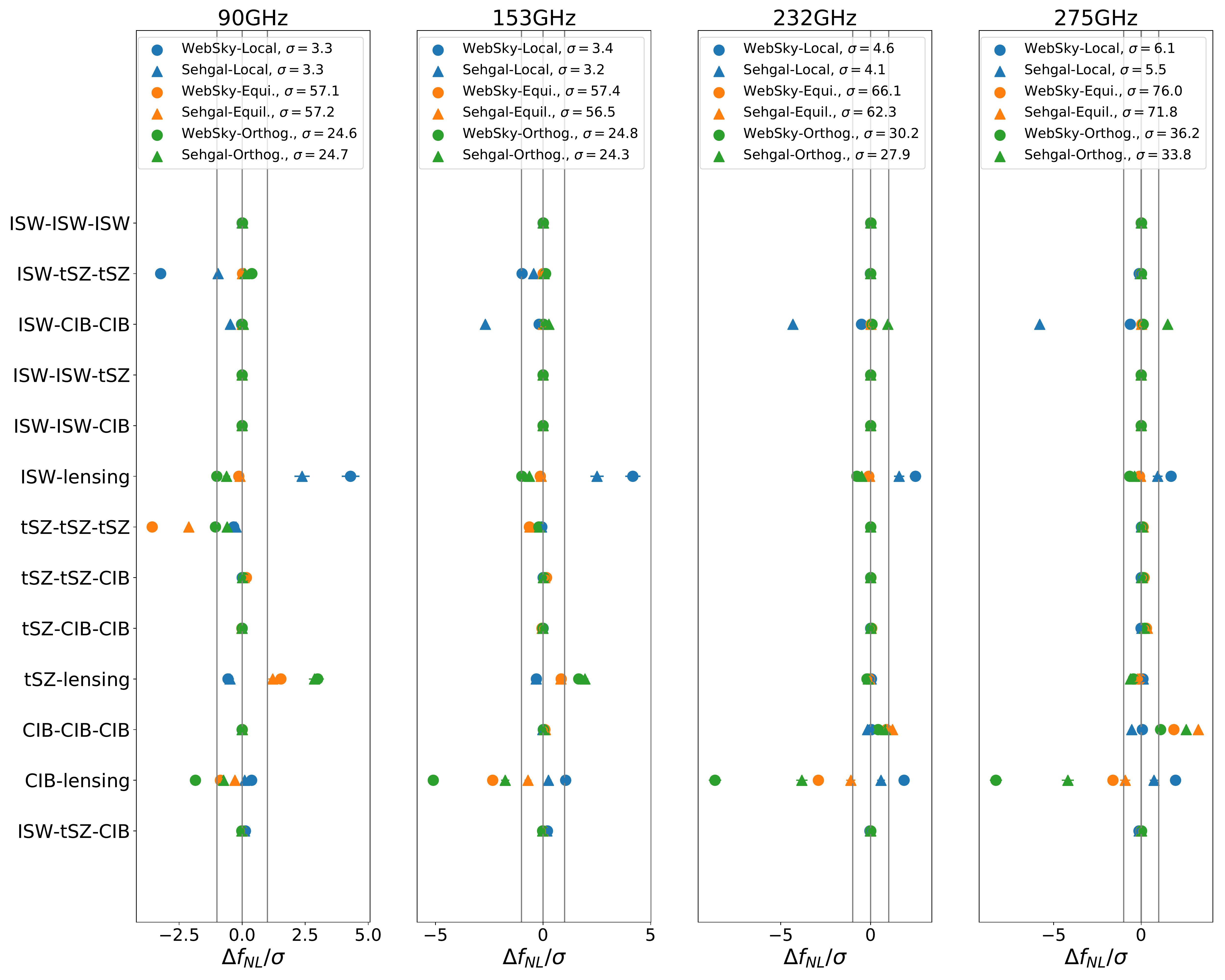}
\caption{Single-frequency biases on $\fnlloc$ (blue), $\fnlequi$ (orange) and $\fnlortho$ (green) from WebSky (circles) and Sehgal et al.~(triangles) simulations for a \textit{Planck}-like (top) and combined \textit{Planck}-like and \textit{SO}-like (bottom) experiments. The biases have been normalised to the 1\,$\sigma$ total error, displayed at the top of each panel for the given primordial template and frequency. The $\fnl$ estimator weights all results by the total sky power including experimental noise, extragalactic foregrounds and the primary CMB power. Note that the error bars represent our uncertainty on the biases from estimating them with a single simulation. They are computed from equation~\eqref{eq:bias_variance} assuming all fields are Gaussian, and thus are likely an underestimate of the full, non-Gaussian errors. The ISW-CIB-CIB local-biases from the Sehgal simulations have been scaled down by a factor of a half for all frequency channels above 200GHz for visualisation purposes. As detailed in Appendix \ref{app:ISW-sehgal}, the large bias is very likely unphyisical, arising from limitations in our approximate ISW map. }
\label{fig:total_bias_plot}
\end{figure*}

In Fig.~\ref{fig:total_bias_plot} we display individual biases from the different CMB secondary anisotropies to the local, equilateral and orthogonal templates relative to the $1\,\sigma$ error for each given experiment (computed via equation~\ref{eq:var_fnl_PNG}) as measured in the Websky and Sehgal et al.~simulations. The total bias, obtained by summing the contributions from all the terms is summarised later in Table~\ref{tab:total_bias_freq}. Due to the inclusion of noise and foregrounds in the variance, the experimental error inflates at the larger frequencies.  The uncertainties on the biases are estimated by equation~\eqref{eq:bias_variance}. A key point to note is that the extragalactic foregrounds are highly non-Gaussian and likely have large connected 4- and 6- point functions \citep{Coulton_2018}. Thus these error bars likely underestimate the true uncertainty in the bias from only having a single simulation. We find that all the radio- and kSZ-related biases are negligible compared to the experimental error and so exclude them from this and subsequent figures for clarity. The radio and kSZ results can be seen in Appendix~\ref{app:RadioAndKSZ}.

In the \textit{Planck} results, the only relevant biases at low frequencies are the lensing biases. At higher frequencies, CIB-related biases begin to dominate as expected due to the strong scaling of the CIB with frequency. While the only significant biasing at 100\,GHz is the ISW-lensing bias, the larger beam size at 100\,GHz for \textit{Planck} inflates the experiment error at this frequency. It is therefore worth performing a multi-frequency analysis in combination with other frequencies rather than attempting to use only the 100\,GHz channel for all PNG estimates. Note that multi-frequency cleaning is also important to suppress biases from Galactic foregrounds \citep{Jung_2018}.

In the SO forecasts, we find at all frequencies several biases are present. At low frequencies for the Websky simulation, the ISW-tSZ-tSZ, tSZ-tSZ-tSZ and tSZ-lensings biases are all (3--4)\,$\sigma$ in the local, equilateral and orthogonal templates, respectively. At higher frequencies, we again find the CIB-related biases to dominate. The CIB-lensing bias is significant in the orthogonal template at all frequencies. These forecasts reinforce how crucial it is for future CMB experiments to use bias-reduction methods, such as multi-frequency foreground cleaning, as single-frequency estimates contain a host of biases. 

For both \textit{Planck} and SO, we see broadly consistent results between the Websky and Sehgal et al.~simulations. As noted earlier, the errors bars we compute are likely underestimated as the non-Gaussian contributions to the variance have been ignored. There is also a second effect at play that complicates the comparison. When computing the biases in the primordial templates, the foreground bispectrum is weighted by the total variance of the map according to equation~\eqref{eq:est_bias}. The variance of the maps is different between the two simulations due to the different levels of foreground power\footnote{This difference is also responsible for the different total experiment errors reported in Fig.~\ref{fig:total_bias_plot} in some cases between the two simulations.} and this leads to a different $\ell$-weighting of the foreground bispectra. For example, consider a hypothetical case where two foregrounds have the same bispectra but one has a larger power spectrum. The larger power spectrum would likely lead to a down-weighting of the foreground bispectrum and hence a smaller bias.  We tested weighting one simulation with the variances from the other. This leads to non-trivial changes in the biases, however these changes do not systematically reduce the difference between the two simulations implying this is not sufficient to explain all the observed differences. The largest difference is seen in the ISW-CIB-CIB bias, however, as is discussed in Appendix~\ref{app:ISW-sehgal}, the method used to construct the ISW field from the Sehgal et al.~simulation is approximate and likely mis-estimates the degree of correlation between the ISW and CIB fields, resulting in an unphysically large bias. The next-largest differences can be seen in the bispectra involving the CIB fields, particularly the CIB-lensing bias. Measurements of the cross power spectrum between the CIB and lensing fields in the two simulations show significant differences, with the websky power spectrum spectrum being generally larger by $\sim 50\%$, and this is the origin of the discrepancy. Note that websky CIB-lensing power spectrum matches well with the measurements from \textit{Planck} that were used in \citet{Hill_2018}, and hence the good agreement between the Websky simulation and \citet{Hill_2018}. This difference is not unexpected as the two simulations use very different modelling of the CIB, as discussed in Section~\ref{sec:simulations}.  Whilst the CIB model in the Websky simulation has been tuned to match the \textit{Planck} observations and thus is expected better to represent the physical bias, there is still a large degree of theoretical uncertainty in modelling the CIB.  Thus a conservative interpretation of the results is that the range of biases produced by the Websky and Sehgal et al.~simulations presents a measure of the theoretical uncertainty on these biases.

\section{Biases from foreground-cleaned CMB maps}\label{sec:CleanedResults}

To mitigate the strong biases discussed above, and those from Galactic emission \citep{Jung_2018}, observations of the sky at a range of frequencies may be combined in a manner that suppresses the foregrounds whilst retaining the CMB signal. In Section~\ref{sec:cleaningMethods} we describe the foreground-cleaning methods considered in this work before presenting the results of these methods applied to a \textit{Planck}-like experiment in Section~\ref{sec:planckForecast_1} and combined \textit{Planck}-like and SO-like experiments in Section~\ref{sec:simonsForecast_1}.

\subsection{Multi-frequency component separation}\label{sec:cleaningMethods}

There exists a large range of possible techniques to combine observations of the sky at different frequencies into a map that contains the CMB anisotropies and minimal other signals (e.g., \citealt{Delabrouille2009}). For example, \citet{planck2016-l04,planck2013-p06,planck2014-a11} detail the application of four component-separation methods that have been used to produce cleaned CMB maps from \textit{Planck} data.

In this work we use the internal-linear-combination (ILC) method \citep{Bennett1992,Tegmark2003,Bennett_2003,Eriksen_2004}. 
This method works by decomposing the set of observations of the sky, $\hat{d}_{\ell m}(\nu_i)$, at frequency $\nu_i$ and multipoles $\ell,m$ into two components: the signal, $s_{\ell m}$, which we want to recover; and the `channel noise' at frequency $\nu_i$, $ n_{\ell m}(\nu_i)$. The signal contributes to an observation at each frequency as $a(\nu_i)s_{\ell m}$, where $a(\nu_i)$ is the spectral response of this component. The ILC method assumes that we know exactly the spectral function and for the CMB this is known to a high precision \citep{COBE_1996}. The channel noise is a broad term that encompasses both experimental noise and those other signals on the sky that are not of interest. The second key assumption of the ILC method is that these other components are uncorrelated with the signal of interest.
The observations may then be written as
\begin{align}
 \hat{d}_{\ell m}(\nu_i) = a(\nu_i)s_{\ell m} + n_{\ell m}(\nu_i) .
\end{align}

The idea of the ILC is to create a linear combination of these maps to recover the signal while minimizing the noise:
\begin{align}
    \hat{s}_{\ell m} = \sum\limits_i w_{\ell m}(\nu_i)\hat{d}_{\ell m}(\nu_i).
\end{align}
This is achieved by optimizing the weights so that they minimize the variance of the recovered signal, with a constraint that the recovered map has unit response to the signal of interest.
This leads to an estimate of the signal as
\begin{align} \label{eq:ILC}
    \hat{s}_{\ell m} = \sum\limits_{i j}\frac{a({\nu_i})[C_\ell^{-1}]^{\nu_i,\nu_j}\hat{d}_{\ell m}(\nu_j) }{a(\nu_i)[C_\ell^{-1}]^{\nu_i,\nu_j} a(\nu_j)},
\end{align} 
where $C_{\ell}^{\nu_i,\nu_j}$ is the cross-power spectrum between frequencies $\nu_i$ and $\nu_j$.

\begin{figure}
\includegraphics[width=0.5\textwidth]{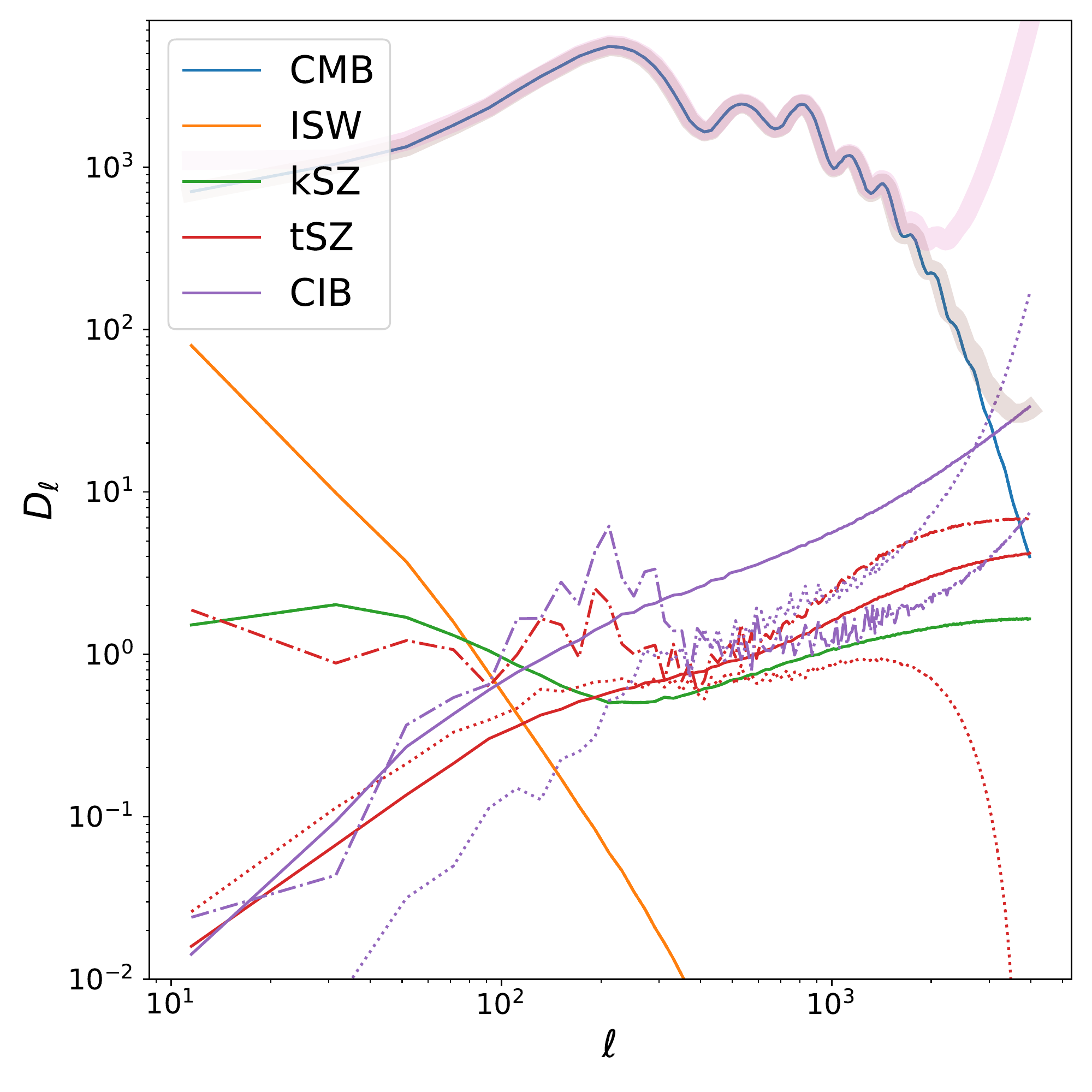}
\caption{Power spectra of the individual sky components at $143$\,GHz (solid lines) and the power spectrum of these components after processing by the \textit{Planck} ILC (dotted lines) and the \textit{Planck}-with-SO ILC (dot-dashed lines). The grey and pink bands denote the total power in the ILC maps for the case of \textit{Planck} only and the combination of \textit{Planck} with SO.}
\label{fig:ilc_ps_components}
\end{figure}
 
The effects of the ILC processing can seen in Fig.~\ref{fig:ilc_ps_components}. We see that the the ILC processing does not necessarily reduce the contribution of all sky signals present at 143\,GHz. The ILC method provides a minimum-variance estimate of the signal of interest, however it puts no constraint on what other signals contribute to the `noise' in the estimated map. The `noise' in the cleaned map will be composed of instrumental noise as well as residual foregrounds, and these foreground signals may still be sufficiently large that they bias inferences of PNG. Another example of large foreground residuals can be seen in \citet{Madhavacheril_2018}, where it is seen that component-separated CMB maps can still contain large contributions from the tSZ effect. We note that the sign of the correlation between the lensing and either tSZ or CIB components on large scales in the \textit{Planck} ILC maps is the same as in the original 143\,GHz map -- thus we should expect the signs of the tSZ-lensing and CIB-lensing biases to be similar to those seen in the 143\,GHz maps. For the ILC map constructed from the combination of 
\textit{Planck} and SO maps, the sign of the tSZ-lensing and CIB-lensing correlations are predominantly the same as at 143\,GHz, but the sign varies across the relevant range of multipoles. %

To ensure that contributions from specific foreground sources are effectively suppressed one may use an extension to the ILC method called the constrained-ILC method, as first developed in \citet{Remazeilles_2011}. This method decomposes observations of the sky as
\begin{align}
 \hat{d}_{\ell m}({\nu_i}) = a({\nu_i})s_{\ell m} +\sum_j b^{(j)}({\nu_i}) x^{(j)}_{\ell m}+n_{\nu_i},
\end{align}
where $b^{(j)}({\nu_i})$ is the known spectral response of the sky contaminant $x^{(j)}_{\ell m}$. We then impose the additional constraint that the estimated map has zero response to these contaminants, i.e.,
\begin{align}
    \sum_i w_{\ell m}({\nu_i}) b^{(j)}({\nu_i}) = 0 \quad \forall j .
\end{align}
We refer to this as deprojecting the relevant sky component(s). For example, if $j=\text{tSZ}$, this would deproject the tSZ signal. See \citet{Madhavacheril_2020} for further details of this method applied to data from the Atacama Cosmology Telescope.

In Sections \ref{sec:planckForecast_1} and \ref{sec:simonsForecast_1} we apply these methods to simulated observations from \textit{Planck} and SO. To apply the ILC method we need the covariance matrix of observations at the different frequency channels, $C_\ell^{\nu_i.\nu_j}$, as is seen in equation~\eqref{eq:ILC}. We compute this by separately measuring the power spectrum of the CMB, extragalactic foregrounds (the Websky or Sehgal et al.~maps) and simulated Galactic foregrounds, obtained from the Python Sky Model \citep{Thorne_2017}. The Galactic foregrounds are measured on $60\,\%$ on the sky that contains the lowest levels of foregrounds  and are computed using the NaMaster pseudo-$C_\ell$ code \citep{Alonso_2019}. To these power spectra we add noise and beam effects based on the experimental configuration. By measuring the components separately, we avoid the `ILC bias'. This is a bias that arises when weighting the maps in the ILC estimate by a covariance matrix estimated from the same maps. See \citet{Delabrouille_2008} for a thorough discussion of the ILC bias.

We note that the ILC method is relatively simple compared to other methods.  However, we expect that the residual biases on scales where the sky is dominated by the CMB or extragalactic sources to be accurately estimated with our ILC method. On these scales the sky signals are statistically isotropic and will be effectively removed by our method. It is less clear how biases from large-scale modes will be modelled. The sky at large scales is more complex: more sky signals contribute to observations (Galactic signals are the dominant contribution) and these show significant spatial variation. State-of-the-art methods account for these complexities and will certainly provide a better suppression of the Galactic foregrounds. However, a priori, it is unclear whether these methods would provide better or worse suppression of large-scale extragalactic components that are still statistically isotropic. Either way, we expect that our ILC method should provide a rough estimate of the level of the PNG biases after foreground cleaning and this estimate is useful to gauge the importance of the possible biases. The important biases can, in future, be assessed with more sophisticated cleaning methods, as necessary.

To complement our ILC method, in Appendix~\ref{app:SMICA_SEVEM} we report results from using two different cleaning methods: approximate versions of the Spectral Matching Independent Component
Analysis (SMICA) and Spectral Estimation Via Expectation
Maximization (SEVEM) component-separation methods. Both these methods were used in the \textit{Planck} non-Gaussianity analyses \citep{planck2016-l04,planck2016-l09}. In the main text, we use the ILC method to maintain a self-contained analysis of the maps (the other two \textit{Planck}-like methods have a set of free parameters that are fixed to the \textit{Planck} data, rather than our simulations and thus may not represent optimal foreground suppression). The results from these other methods are very similar to the ILC results reported below.

\subsection{Biases to \textit{Planck}-like experiments}\label{sec:planckForecast_1}

In this analysis, we consider observations at $30$\,GHz, $44$\,GHz, $70$\,GHz, $100$\,GHz, $143$\,GHz, $217$\,GHz, $353$\,GHz and $545$\,GHz with noise and beam properties as reported in \citet{planck2016-l01}.
Note that when considering the Sehgal et al.~maps our ILC excludes the $545$\,GHz channel as there are no CIB maps provided for this frequency. \footnote{We do not consider the $857$\,GHz channel as the calibration uncertainty on the \textit{Planck} measurements at $857$\,GHz is large, which means it challenging to include this frequency in component separation pipelines. } As the kSZ and radio biases remain small from the ILC-cleaned maps, we defer further discussion of them to Appendix~\ref{app:RadioAndKSZ}.  In addition to point source mask, discussed in Section \ref{sec:simulations}, we include a mask on the brightest clusters that removes clusters with masses above $M\geq 5 \times 10^{14} M_{\sun}h^{-1}$. 

\begin{figure*}
\includegraphics[height=13cm,width=16cm]{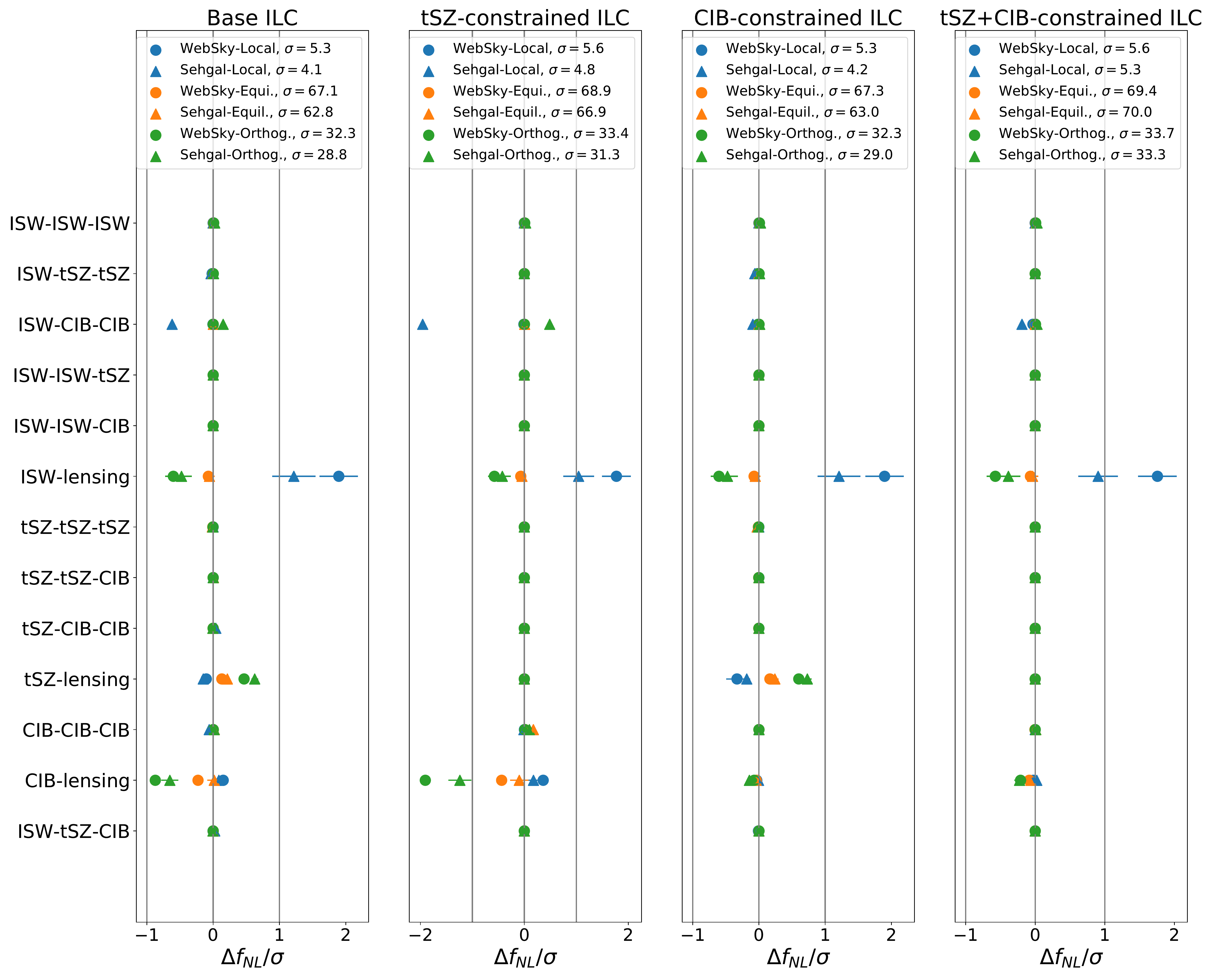}
\caption{Biases to PNG estimated from the WebSky (circles) and Sehgal et al.~(triangles) simulations after ILC foreground cleaning for a \textit{Planck}-like experiment. Results are normalised by the 1\,$\sigma$ experiment error (displayed in the legend of each panel). Almost all biases are below the experiment error after ILC cleaning, with only the tSZ-lensing and/or CIB-lensing biases remaining significant; the un-normalised results of these biases are shown in Table~\ref{tab:Planck_ILC_specific_biases}. The constrained ILCs inflate the experiment error, with the largest inflation coming from constraining the tSZ signal. The error bars are computed as in Fig.~\ref{fig:total_bias_plot}.  As in Fig.  \ref{fig:total_bias_plot}, the unphysically large ISW-CIB-CIB local bias from the Sehgal maps for the Base ILC and tSZ-constrained ILC has again been scaled down by a factor of a half . }
\label{fig:Planck_ILC_biases}
\end{figure*}

The biases from each component after multi-frequency cleaning of the Websky and Sehgal et al.~maps are shown in Fig.~\ref{fig:Planck_ILC_biases} and the total biases are shown later in Table~\ref{tab:total_bias_ILC}. Four sets of biases are shown corresponding to the following ILCs: a base ILC without any further constraints; an ILC that deprojects the tSZ signal only; an ILC that deprojects the CIB signal only; and an ILC that deprojects both the tSZ and CIB signals. The biases shown are normalised to the experiment 1\,$\sigma$ error of the corresponding primordial template and ILC, as calculated by evaluating equation~\eqref{eq:var_fnl_PNG} with the total power spectrum computed from the ILC. Note that, as expected, the constrained-ILC methods produce larger experiment errors, with the largest increase arising from constraining the tSZ signal (this arises as the large CIB signal at high frequencies makes it simpler to be completely removed compared to the tSZ signal, which is subdominant at all frequencies).

\begin{table*}
 \begin{tabular}{c c c c c c} 
 \hline
Bias type &  ILC type & bias to $\fnlloc$ & bias to $\fnlequi$ & bias to $\fnlortho$ & bias to $\fnliswlens$ \\
 \hline  
\multirow{4}{*}{CIB-lensing} & Base & $0.80\pm0.24$ & $-15\pm3.9$ & $-28\pm2.2$ & $0.10\pm0.01$  \\  
& tSZ deprojected & $2.0\pm0.39$ & $-30\pm6.2$ & $-64\pm3.6$ & $0.23\pm0.03$  \\  
& CIB deprojected  & $-0.17\pm0.21$ & $-2.8\pm3.0$ & $-2.3\pm1.8$ & $0.01\pm0.01$ \\  
& tSZ \& CIB deprojected  & $-0.14\pm0.35$ & $-6.0\pm5.2$ & $-7.1\pm3.1$ & $0.02\pm0.02$ \\  
 \hline  
\multirow{4}{*}{tSZ-lensing} &  Base & $-0.55\pm0.27$ & $8.8\pm2.2$ & $15\pm1.3$ & $-0.10\pm0.02$  \\
& tSZ deprojected  & 0.00 & 0.00 & 0.00 & 0.00 \\  
& CIB deprojected   & $-1.7\pm0.86$ & $11\pm3.1$ & $20\pm2.5$ & $-0.11\pm0.02$ \\ 
& tSZ \& CIB deprojected   & 0.00 & 0.00 & 0.00 & 0.00 \\  
 \hline
 \end{tabular}
 \caption{Biases to the local, equilateral and orthogonal types of PNG for a \textit{Planck}-like experiment from CIB-lensing and tSZ-lensing bispectra, after component separation, in the WebSky ILC maps. The errors presented in the table are the Gaussian errors for each specific bias. The 1\,$\sigma$ total experiment error is much larger, with the values for the base ILC being: $\sigma(\fnlloc)=5.27$; $\sigma(\fnlequi)=67.1$; and $\sigma(\fnlortho)=32.3$. Relative to the experiment errors, the local and equilateral biases are not significant, but the orthogonal biases are almost at the 1\,$\sigma$ level for the standard ILC. Constraining either the tSZ or CIB individually suppresses the respective bias, but inflates the other bias. We also include biases to an estimate of the amplitude of the ISW-lensing bispectrum, with fiducial value $\fnliswlens=1$.
 }
\label{tab:Planck_ILC_specific_biases}
\end{table*}

We find that several biases present in the single-frequency maps are either suppressed or completely removed by just the standard ILC. The remaining frequency-dependent biases -- CIB-lensing and tSZ-lensing -- can be almost completely removed by the appropriate constrained ILC.\footnote{Deprojecting the CIB assuming a specific scaling with frequency does not completely remove the CIB since the emission decorrelates slightly across frequencies.} In Table~\ref{tab:Planck_ILC_specific_biases} we show the un-normalised biases to PNG for these specific contaminating bispectra. The bias from these bispectra is most significant in the orthogonal limit. 
In particular, the CIB-lensing $\fnlortho$ bias is around the $1\,\sigma$ level for both the standard and tSZ-constrained ILC, whereas the tSZ-lensing $\fnlortho$ bias is around the $1\,\sigma$ level for both the standard and CIB-constrained ILC. While the presence of these biases in our results suggests possible contamination of the results from \textit{Planck}, it should be noted that these biases are of equal magnitude but opposite signs, and so the overall bias for the standard ILC is likely to be small, as is seen in Table~\ref{tab:total_bias_ILC}. Note that when we deproject only the CIB or tSZ, the \emph{total} bias from the correlation of these components with lensing grows significantly over the case of the standard ILC.
This is for two reasons: firstly, there is no cancellation between the two sources of bias; and, secondly, the weights that minimise the variance while deprojecting one of these foregrounds typically result in larger residuals of the non-deprojected component. This is seen in Fig.~\ref{fig:Planck_ILC_biases} where the CIB-lensing bias grows when we deproject the tSZ effect. The only bias that is independent of frequency is the ISW-lensing bias. As this can be very accurately calculated, removing this bias is not a significant issue. 

\begin{table*}
 \begin{tabular}{c c c c c c}
 \hline
 Bias type & ILC type & bias to $\fnlloc$ & bias to$\fnlequi$ & bias to $\fnlortho$ & bias to $\fnliswlens$ \\
 \hline  
\multirow{4}{*}{CIB-lensing}& Base & $1.0\pm0.24$ & $-46\pm 4.6$ & $-47\pm2.4$ & $0.21\pm0.02$  \\  
& tSZ deprojected   & $1.6\pm0.37$ & $-57\pm6.5$ & $-76\pm3.5$ & $0.34\pm0.03$ \\  
& CIB deprojected    & $-0.23\pm0.13$ & $-2.25\pm2.35$ & $-2.4\pm1.2$ & $-0.01\pm0.01$ \\  
& tSZ \& CIB deprojected    & $-0.37\pm0.33$ & $-8.1\pm5.5$ & $-7.8\pm3.1$ & $0.00\pm0.03$ \\  
 \hline  
\multirow{4}{*}{tSZ-lensing} &  Base & $-1.5\pm0.40$ & $56\pm7.2$ & $39\pm3.7$ & $-0.29\pm0.03$  \\  
& tSZ deprojected   & 0.00 & 0.00 & 0.00 & 0.00 \\  
& CIB deprojected    & $-1.5\pm0.40$ & $58\pm7.3$ & $43\pm3.8$ & $-0.29\pm0.03$ \\  
& tSZ \& CIB deprojected   & 0.00 & 0.00 & 0.00 & 0.00 \\  
 \hline
 \end{tabular}
 \caption{Biases to the local, equilateral and orthogonal types of PNG for the combination of \textit{Planck}-like and SO-like experiments from CIB-lensing and tSZ-lensing bispectra, after component separation, in the WebSky ILC maps. The setup is otherwise identical to Table~\ref{tab:Planck_ILC_specific_biases}.  The 1\,$\sigma$ total experiment error for the base ILC being is: $\sigma(\fnlloc)=3.3$; $\sigma(\fnlequi)=57$; and $\sigma(\fnlortho)=25$. }
\label{tab:SO_ILC_specific_biases}
\end{table*}

\begin{figure*}
\includegraphics[height=13cm,width=16cm]{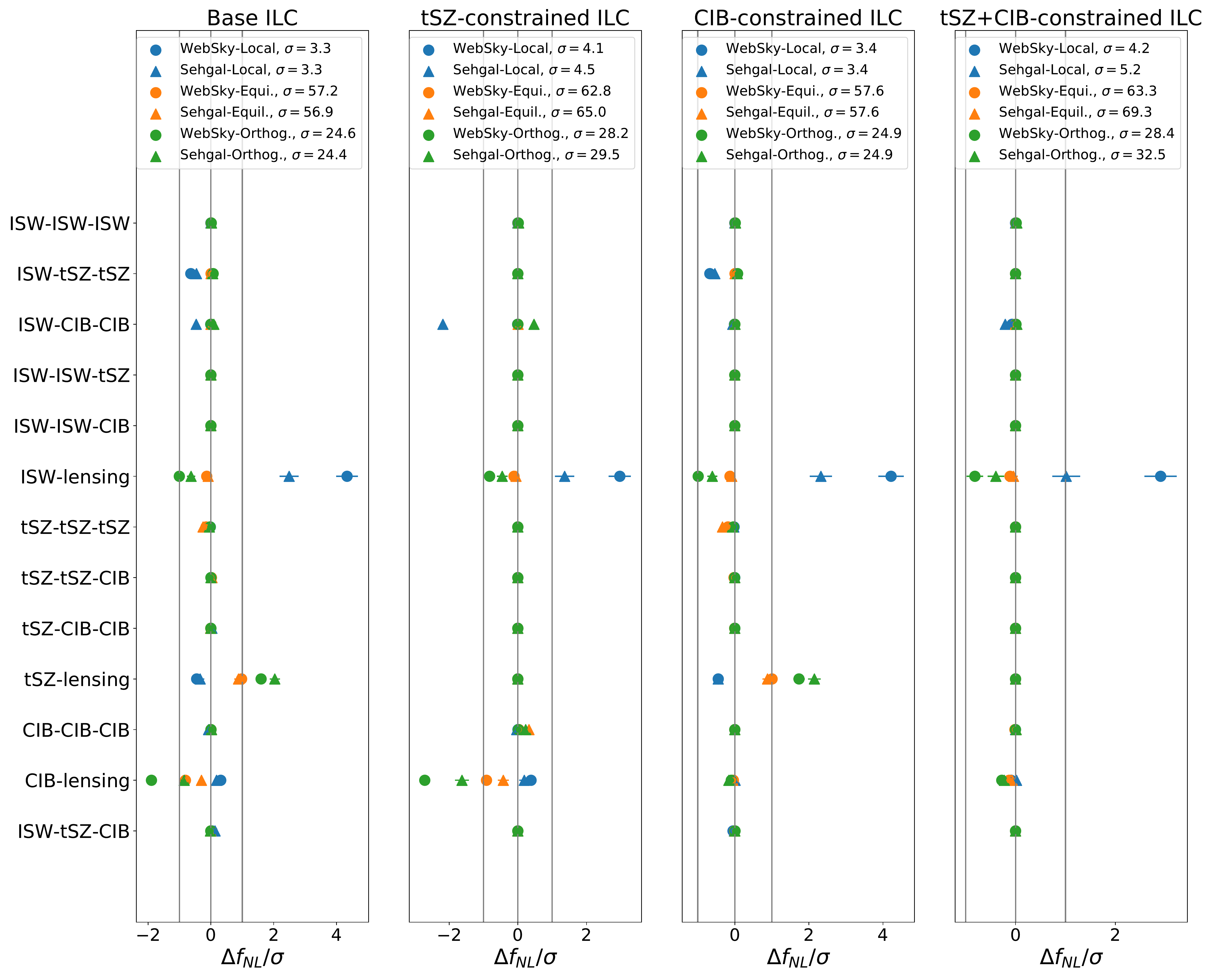}
\caption{Biases to PNG estimated from the WebSky (circles) and Sehgal et al.~(triangles) simulations after ILC foreground cleaning for the combination of \textit{Planck}-like and SO-like experiments. Results are normalised by the 1\,$\sigma$ experiment error (displayed in the legend of each panel). Otherwise, the specification is the same as Fig.~\ref{fig:Planck_ILC_biases}, including the rescaling of the ISW-CIB-CIB local bias from the Sehgal maps.}
\label{fig:SO_ILC_biases}
\end{figure*}

We generally see very consistent results between the Sehgal et al.~and Websky simulations. However, there is one anomalous result: the seemingly large ISW-CIB-CIB local-type bias from the Sehgal et al.~maps. As described in Section~\ref{sec:total_bias}, this is likely not physical and thus not important.

Given the non-trivial biases from lensing-induced bispectra, it is interesting to investigate the bias on measurements of the ISW-lensing bispectrum. As shown in \citet{planck2013-p09a}, the value of the ISW-lensing bispectrum measured by \textit{Planck} is lower than its expected value.
It is interesting to investigate whether this could be due to foreground biases from other lensing-induced bispectra. We quantify its strength via $\fnliswlens$, computed as in \citet{Lewis_2011}. As opposed to the other templates, it is theoretically expected for the strength of this shape to be unity, i.e., $\fnliswlens=1$. The biases from the CIB-lensing and tSZ-lensing bispectra to this shape are included in Table \ref{tab:Planck_ILC_specific_biases}. For the base ILC, we find that the contributions of the CIB-lensing and tSZ-lensing bispectra together only shift the value of $\fnliswlens$ by $\Delta\fnliswlens=-0.002$. This small shift occurs due to a cancellation in the two sources of bias, making the net bias negligible and far below experimental error ($\sigma(\fnliswlens)=0.25$). 
We also note that specifically deprojecting either one of the CIB or tSZ in the ILC leads to larger biases in the value of $\fnliswlens$. Such shifts were seen in the \textit{Planck} analysis~\citep{planck2016-l09}, although it was reasoned to arise from the vanishing of the tSZ-lensing and ISW-tSZ-tSZ bispectra (when deprojecting the tSZ) such that the remaining signal was just the ISW-lensing bispectrum. Our results suggest this shift occurred due to the CIB-lensing bispectrum no longer cancelling with the tSZ-lensing bispectrum, and so the somewhat-low results from \textit{Planck} \emph{without} deprojection may be a more accurate measurement of the ISW-lensing signal for our sky. Although this conclusion may be sensitive to the precise details of the analysis. 

In summary, we find no large biases to PNG measurements from a \textit{Planck}-like experiment after a simple ILC foreground cleaning. The largest deviations would be in the $\fnlortho$ bias, but even these deviations are around one-third of the experimental error. We note that this is not because the standard ILC removes all the foregrounds but from a cancellation of two biases, each individually close to the $1\,\sigma$ level.

\subsection{Biases to Simons Observatory-like experiments}\label{sec:simonsForecast_1}

In this analyses we consider combining the \textit{Planck}-like observations of the previous section with SO-like observations at $27$\,GHz, $42$\,GHz, $90$\,GHz, $153$\,GHz, $232$\,GHz and $275$\,GHz with instrumental noise properties from \citet{SO_2019}. We use the atmospheric noise model from  \citet{SO_2019} to include the large atmospheric noise component. A point-source mask with a 15\,mJ flux-density cut at $217$\,GHz has again been applied to mimic experimental procedures, where bright sources from the CIB \footnote{The radio maps we use already have a 7\,mJ flux-density cut applied to the}, would otherwise obscure results (see Appendix \ref{app:PSmasking} for a detailed discussion). We also mask galaxy clusters with masses above $M\geq 5 \times 10^{14} M_{\sun}h^{-1}$. This represents an approximate level of clusters that can be detected and removed for SO and will reduce the biases from the tSZ effect. Note that, again, when considering the Sehgal et al.~maps our ILC excludes the $545$\,GHz maps as there are no CIB maps provided for this frequency.

 Our forecast biases from each component after foreground cleaning are shown in Fig.~\ref{fig:SO_ILC_biases} and the total biases are shown later in Table~\ref{tab:total_bias_ILC}. We consider the same four different ILC types and display results from the Sehgal et al.~and Websky simulations. As before, the biases shown in the figure are normalised to the $1\,\sigma$ experiment error and the kSZ and radio biases, which are small, are presented in Appendix~\ref{app:RadioAndKSZ}.

These forecasts combining \textit{Planck}-like and SO-like experiments are largely consistent with our previous \textit{Planck}-only results. The majority of biases seen at single frequencies now vanish. The CIB-lensing and tSZ-lensing bispectra still have a strong signal in the orthogonal limit, but the biases are opposite and so largely cancel each other out. Similar to the \textit{Planck}-like results, deprojecting just one component significantly increases the bias from the other lensing source. 

Our forecasts show additional biases, at non-neglibile levels, may be present in the \textit{Planck}-like plus SO-like combination. In particular, we find that the ISW-tSZ-tSZ bispectrum $\fnlloc$ bias is present at close to the $1\,\sigma$ experiment error level in the standard and CIB-deprojected ILCs. This bias was seen to be strong in the single-frequency measurements (Table~\ref{tab:compare_ISW}) and has not been completely removed with foreground cleaning. This would have to be accounted for via marginalisation or subtraction, or alternatively, using the tSZ-constrained ILC. As the cost of additionally deprojecting the CIB is low in comparison to the tSZ, it is recommended always to deproject \emph{both} the tSZ and CIB over only the tSZ. This also eliminates the contamination from the lensing biases that we further discuss below. Note that if we do not apply the tSZ cluster mask, the ISW-tSZ-tSZ and tSZ-tSZ-tSZ biases are 2--4 times larger (see Appendix~\ref{app:clusterMasking}).

As before the only frequency-independent bias is the ISW-lensing bias, which again can be accurately modelled and removed.

As also in the \textit{Planck} case, we generally see very good agreement between the Sehgal et al.~and the Websky simulations. The exception is the unphysical ISW-CIB-CIB bias in the Sehgal et al. simulation. 
This is interesting as our analysis of the Sehgal et al.~simulations uses a smaller range of frequencies (only up to 353\,GHz). This suggests that, at the level of biases relevant for upcoming experiments, the CIB can be sufficiently mitigated just with observations up to 353\,GHz.

As before, we consider the biases from lensing bispectra to measurements of the ISW-lensing bispectrum, quantified through $\fnliswlens$. The results of this are shown in Table~\ref{tab:SO_ILC_specific_biases} along with the biases to the three primordial shapes. Summing over these lensing biases, we find that the extra CIB-lensing and tSZ-lensing signals bias $\fnliswlens$ by $\Delta\fnliswlens=-0.077$ for the base ILC. Deprojecting the tSZ effect changes the total lensing bias to $\Delta\fnliswlens=+0.342$ and deprojecting the CIB changes it to $\Delta\fnliswlens=-0.302$. Deprojecting both signals changes the bias to $\Delta\fnliswlens=+0.004$. The least overall lensing bias comes from deprojecting both components,
but the base ILC does better than any individually constrained ILC.
\section{Conclusion}\label{sec:conclusion}

We have considered how secondary anisotropies in the CMB bias measurements of primordial non-Gaussianity. Motivated by the work of \citet{Hill_2018}, we performed an extensive analysis of possible biases arising from all CMB secondary anisotropies: the ISW, tSZ and kSZ effects, the CIB, radio galaxies and gravitational lensing. We used the binned bispectrum estimator to measure all the biases present in the full-sky Websky and Sehgal et al.~simulations. We cross-checked our results with the analytical, cosmic-variance-limited biases studied in \citet{Hill_2018}, namely lensing and ISW-related biases, to validate our approach and found generally good agreement. We then included experimental effects (noise, beam and foregrounds) in our estimator to mimic better experimental conditions for single-frequency measurements.

\begin{table*}
 \begin{tabular}{c c c c c c c c c}
 \hline
\multirow{2}{*}{Experiment} & \multirow{2}{*}{Frequency (GHz)} &  \multicolumn{2}{c}{$\Delta\fnlloc$} & \multicolumn{2}{c}{$\Delta\fnlequi$} & \multicolumn{2}{c}{$\Delta\fnlortho$} \\
&& Total bias & $\sigma_{\mathrm{Exp.}}$ & Total bias & $\sigma_{\mathrm{Exp.}}$ & Total bias & $\sigma_{\mathrm{Exp.}}$ \\
 \hline

\multirow{4}{*}{\textit{Planck}} & 100 &  $-0.8\pm0.4$ & $7.8$ & $-16\pm5.2$ & $81$ & $-2.4\pm3.2$ & $41$ \\  
& 143 & $0.3\pm0.4$ & $5.5$ & $-21\pm6.0$ & $70$ & $-30\pm3.4$ & $34$ \\  
& 217  & $4.7\pm0.9$ & $5.5$ & $-37\pm13$ & $70$ & $-140\pm6.9$ & $34$\\  
& 353  & $11\pm6.3$ & $13$ & $890\pm51$ & $110$ & $43\pm22$ & $55$ \\  
 \hline

\multirow{4}{*}{SO} &  90 & $-11\pm0.7$ & $3.3$ & $-160\pm13$ & $57$ & $7.2\pm6.7$ & $25$\\
& 145  & $-0.2\pm0.9$ & $3.4$ & $-120\pm14$ & $57$ & $-89\pm7.4$ & $25$\\  
& 232   & $5.9\pm1.6$ & $4.6$ & $-140\pm21$ & $66$ & $-260\pm12$ & $30$\\ 
& 275   & $6.9\pm2.3$ & $6.1$ & $62\pm26$ & $76$ & $-270\pm15$ & $36$\\  
 \hline
 \end{tabular}
 \caption{Total biases in the local, equilateral and orthogonal primordial shapes for a \textit{Planck}-like or SO-like experiment from all bispectra at individual frequencies in the WebSky maps. Here, $\sigma_{\mathrm{exp}}$ represents the experimental error whereas the data-point errors represent the Gaussian errors of the total bias as measured in the simulations (and so approximate our uncertainty on the bias from having estimated it from a single simulation). Biases that are comparable to the experimental error are a potential concern for future experiments.  These biases do not include the ISW-lensing bias, which is well known and easily calculated; see, e.g.,~\citet{Lewis_2011}.} 
\label{tab:total_bias_freq}
\end{table*}

\begin{table*}
 \begin{tabular}{c c c c c c c c c}
 \hline
\multirow{2}{*}{Experiment} & \multirow{2}{*}{Frequency (GHz)} &  \multicolumn{2}{c}{$\Delta\fnlloc$} & \multicolumn{2}{c}{$\Delta\fnlequi$} & \multicolumn{2}{c}{$\Delta\fnlortho$} \\
&& Total bias & $\sigma_{\mathrm{Exp.}}$ & Total bias & $\sigma_{\mathrm{Exp.}}$ & Total bias & $\sigma_{\mathrm{Exp.}}$ \\
 \hline

\multirow{4}{*}{\textit{Planck}} & Base &  $0.1\pm0.4$ & $5.3$ & $-6.6\pm4.5$ & $67$ & $-13\pm2.6$ & $32$\\  
& tSZ deprojected & $2.0\pm0.4$ & $5.6$ & $-29\pm6.2$ & $69$ &  $-63\pm3.6$ & $33$\\  
& CIB deprojected  & $-2.15\pm0.9$ & $5.3$ & $7.8\pm4.3$ & $67$ & $17\pm3.1$ & $32$\\  
& tSZ \& CIB deprojected  & $-0.4\pm0.4$ & $5.6$ & $-5.8\pm5.2$ & $69$ & $-6.7\pm3.1$ & $34$\\  
 \hline  
 
\multirow{4}{*}{SO} &  Base & $-2.5\pm0.5$ & $3.3$ & $2.3\pm8.6$ & $57$ & $-5.9\pm4.5$  & $25$\\
& tSZ deprojected  & $1.6\pm0.4$ & $4.1$ & $-55\pm6.5$ & $63$ & $-75\pm3.6$ & $28$ \\  
& CIB deprojected   & $-4.3\pm0.4$ & $3.4$ & $44\pm7.7$ & $58$ & $42\pm4.1$ & $25$ \\ 
& tSZ \& CIB deprojected   & $-0.7\pm0.3$ & $4.2$ & $-8.3\pm5.6$ & $63$ & $-7.3\pm3.2$ & $28$      \\  
 \hline
 \end{tabular}
 \caption{Total biases in the local, equilateral and orthogonal primordial shapes, for a \textit{Planck}-like experiment alone or in combination with an SO-like experiment, from all bispectra after component separation in the WebSky ILC maps. As in Table~\ref{tab:total_bias_freq}, $\sigma_{\mathrm{exp}}$ represents the experimental error and the data point errors are the Gaussian errors as measured from the simulations. The ISW-lensing bias is again subtracted from these results.}
\label{tab:total_bias_ILC}
\end{table*}

We find that a range of biases exist for single-frequency measurements in the \textit{Planck} HFI channels of 100, 143, 217 and 353\,GHz with the strength and number of biases only growing for a future SO-like experiment. As expected due to the frequency dependence of the signals, we find that tSZ-related biases dominate the low-frequency maps and CIB-related biases dominate the higher frequencies. The total biases at each frequency to the amplitudes $\fnlloc$, $\fnlequi$ and $\fnlortho$ of primordial non-Gaussianity of local, equilateral and orthogonal type for the \textit{Planck}-like and SO-like experiments are shown in Table \ref{tab:total_bias_freq}.

The frequency dependence of the tSZ and CIB differs from the CMB fluctuations, so that combinations of the different frequency maps can be used in an attempt to remove these foregrounds. We considered the effectiveness of one foreground cleaning method, the internal-linear-combination method, on reducing the biases arising from bispectra involving the tSZ and CIB. We find that for \textit{Planck}, the only frequency-dependent biases that remain after a standard ILC are the CIB-lensing and tSZ-lensing biases with the most significant biases being for $\fnlortho$. However, we find that these biases are of similar strength but opposite sign and so their overall impact is likely small. This is seen in Table \ref{tab:total_bias_ILC} where the total biases after the base ILC are found to be smaller than the experimental error. Constraining for one particular foreground in the ILC only removes this cancellation and so it is recommended that either both foregrounds, or neither, should be constrained in the ILC.

Combining \textit{Planck}-like and SO-like measurements in the ILC, we see the same effect from the lensing biases, with greater than $1\,\sigma$ biases from each but the overall bias is approximately $\sigma/3$. In addition, we also find that biases from the tSZ-tSZ-tSZ and ISW-tSZ-tSZ bispectra are present in the standard ILC results. These are completely removed when constraining for the tSZ effect in the ILC.  Whilst the estimates of the total biases, shown in Table \ref{tab:total_bias_ILC}, are small, these arise from the cancellation of multiple larger biases. The degree of cancellation is likely sensitive to the modelling and the exact cleaning procedure, thus to minimize possible contamination it is likely that constraining \emph{both} the tSZ and CIB signals int the foreground cleaning  will be required.

Throughout, we found broadly consistent results between the Websky and Sehgal et al.~simulations. The two simulations allow us to estimate roughly the degree of systematic modelling uncertainty on our inferred biases. However this is still a limited estimate. A more robust estimation of the possible biases would need larger suites of simulations with multiple realisations of the foreground components. 

We considered component-separation methods to remove these biases. However these methods do come at the cost of decreased constraining power -- particularly when deprojecting the tSZ signal. In future it would be useful to investigate alternative methods to reduce the biases. One such method would be fitting for foreground templates jointly with the primordial templates. A second method could be to use CMB delensing to reduce all of the lensing-related biases. Delensing methods have already been extensively developed, for removing biases to primordial B-mode searches \citep{Knox_2002,Kesden_2002,Seljak_2004,Smith_2012,sherwin_2015,Adachi_2020} and sharpening the CMB peaks \citep{Carron:2017vfg,Sehgal_2017,Green_2017}, and thus can be easily incorporated into PNG analyses.  Delensing has the further benefit, as was shown in \citet{Coulton_2020}, that it can reduce the variance on bispectrum estimators.

\section*{Acknowledgements}
The authors would like to thank Daan Meerburg, David Spergel, Paul Shellard, James Fergusson and Colin Hill for useful discussions.
AC acknowledges support from the STFC (grant numbers ST/N000927/1 and ST/S000623/1). This research used resources of the National Energy Research Scientific Computing Center (NERSC), a U.S. Department of Energy Office of Science User Facility.  The Flatiron Institute is supported by the Simons Foundation. AM is supported by the STFC and the EPSRC funded centre for doctoral training. This work made use of the \textsc{pixell} library\footnote{\url{https://github.com/simonsobs/pixell}}, \textsc{numpy} \citep{harris2020array},\textsc{scipy} \citep{2020SciPy-NMeth} and \textsc{matplotlib} \citep{Hunter:2007}
\section*{Data Availability}
The Websky simulations are available at: \url{https://mocks.cita.utoronto.ca/data/websky/v0.0/} and the Sehgal et al simulations are available at \url{https://lambda.gsfc.nasa.gov/simulation/full_sky_sims_ov.html}. Additionally we use data available within the PySM library available at: \url{https://pysm3.readthedocs.io/en/latest/}.
\appendix
\section{Generating integrated Sachs--Wolfe maps for the Sehgal et al.~simulations}\label{app:ISW-sehgal}
The integrated Sachs--Wolfe field is linear on large scales, which means that it can be simulated with high accuracy by drawing from a Gaussian random field. Given a realization of other extragalactic fields we can generate a correlated realization of the ISW effect using the covariance of the ISW effect with the other CMB secondaries. 

Specifically, we use the following procedure to generate an approximate ISW map for the Sehgal et al.~simulation. For each of the Websky and Sehgal et al.~simulations, we measure the power spectra $\Sigma_\ell^{ij}$
between all the secondary anisotropies.
For the Sehgal et al.~simulations, $\Sigma_\ell$ is a $3\times 3$ matrix 
as we compute the correlations between the lensing convergence, CIB and tSZ maps, whilst for the Websky maps it is $4\times 4$ as we also include the correlations with the ISW. For notational clarity, we shall sometimes explicitly denote the dimension of the covariance matrices, i.e., $\Sigma_\ell^{(3)}$ and  $\Sigma_\ell^{(4)}$. We then create a set of unit-variance complex random variables, $g^i_{\ell m}$, from the Sehgal et al.~simulations' spherical harmonic coefficients as
\begin{align}
   g^i_{\ell m} = \sum\limits_{j=1}^{3}\left[\left(\Sigma_\ell^{(3)\,\text{Sehgal}}\right)^{-1/2}\right]^{ij} a^{j,\mathrm{Sehgal}}_{\ell m}.
\end{align}
Here, the inverse square root of the Sehgal et al.~power spectrum covariance is taken with a Cholesky decomposition and is a lower-triangular matrix.
This vector is augmented with a set of randomly drawn unit-variance complex numbers so that we extend the $i$ index by one. This additional component will be the random part of the new ISW field that is uncorrelated with the tSZ, CIB and lensing convergence.
If the cross-power spectra of the Websky and Sehgal et al.~simulations were equal for the tSZ, CIB and lensing convergence fields, we could generate an appropriately correlated ISW field for the Sehgal et al.~simulation as
\begin{equation}
    \hat{a}^{\rm{ISW}}_{\ell m} =\sum\limits_{j=1}^{4} \left[\Sigma^{1/2}_\ell\right]^{\text{ISW} j} g^j_{\ell m},
\end{equation}
where the $4\times 4$ covariance matrix here is from Websky. In practice, there are large differences in the power spectra of the two simulations on large scales, possibly due to the fact that the Sehgal et al.~simulations are constructed from one octant that is then replicated. We deal with this by using the correlation structure from Websky, but retaining the tSZ, CIB and lensing convergence auto-power spectra from Sehgal et al.~(and taking the ISW power from Websky). This means that
we rescale the elements of the Websky covariance matrix by
\begin{align}
   \left[\tilde{\Sigma}_\ell^{(4)}\right]^{ij} =  \sqrt{\frac{ C^{ii, \mathrm{Sehgal}} C^{jj, \mathrm{Sehgal}} }{C^{ii, \mathrm{Websky}}C^{jj, \mathrm{Websky}}} } \left[\Sigma_\ell^{(4)\,\text{Websky}}\right]^{ij} .
\end{align}
Finally, we generate the Sehgal et al.~ISW map using this rescaled covariance matrix as
\begin{equation}
    \hat{a}^{\rm{ISW}}_{\ell m} =\sum\limits_{j=1}^{4} \left[\tilde{\Sigma}^{1/2}_\ell\right]^{\text{ISW} j} g^j_{\ell m}.
\end{equation}

This procedure can be analogously understood as a minimum-variance reconstruction. In this view, we want to reconstruct an ISW map, $\hat{a}^\mathrm{ISW}_{\ell m}$, as a linear combination of the three foreground maps, i.e.,
\begin{align}
    \hat{a}^\mathrm{ISW}_{\ell m} = \sum\limits_{i=1}^{3} \gamma^i_{\ell m}\, a^{i}_{\ell m} + u_{\ell m},
\end{align}
where the $\gamma^i_{\ell m}$ give the contributions from each correlated component and $u_{\ell m}$ is a vector of Gaussian random variables, which accounts for the uncorrelated component. We require that the cross-correlation of this reconstructed field with any of the other components gives the observed power spectrum, i.e.,
\begin{align}
    \langle \hat{a}^\mathrm{ISW}_{\ell m}  \left(a^i_{\ell m}\right)^\ast \rangle = \gamma^i_{\ell m} C^{ii}_{\ell} = C_\ell^{\mathrm{ISW}\,i},
\end{align}
and itself has the correct ISW power spectrum.
In our case, the cross-spectra, $C_\ell^{\mathrm{ISW}i}$, are rescaled versions measured from the Websky simulations. Given these constraints, we find the minimum-variance solutions for the coefficients, $\gamma^i_{\ell m}$. Note these methods are only exactly equivalent if the matrix square roots in the first method are lower triangular computed using the Cholesky decomposition (which is what we used).

As we perform this operation on 2D fields that are projected along the line-of-sight, as opposed to on the 3D fields that are subsequently projected, this method will generate only an approximate ISW map. It should be useful to estimate the impact of ISW related non-Gaussianites but should be treated only as an estimate. The redshift distribution of the perturbations that contribute to our simulated ISW map is unlikely to match the true redshift distribution (a product of generating a 2D map) and so may lead to unphysically large, or small, non-Gaussianties. This is observed in our results for the ISW-CIB-CIB bias such as in Fig.~\ref{fig:total_bias_plot}. 

\section{SMICA and SEVEM results}\label{app:SMICA_SEVEM}
In Section \ref{sec:CleanedResults} we presented an estimate of the biases present in a \textit{Planck}-like experiment. The \textit{Planck} collaboration used a wide range of component-separation methods in their primordial non-Gaussianity analyses, as detailed most recently in \citet{planck2016-l09}, and to assess accurately any biases in those results it would be necessary to apply the same component-separation methods to our simulations. The techniques used there are, generally more sophisticated than our ILC method and so it is possible that the level of biases are further suppressed. In this appendix we present an estimate of the level of the biases in the data after applying approximations of the Spectral Estimation Via Expectation
Maximization (SEVEM)  and Spectral Matching Independent Component
Analysis (SMICA) methods used in the \textit{Planck} analyses \cite[see, e.g.,][for more details]{planck2013-p06,planck2016-l04}. These methods require observations at all \textit{Planck} frequencies. For the Sehgal et al.~simulations, we generate high-frequency CIB maps at $\nu>353$\,GHz by rescaling the 353\,GHz map using a modified blackbody spectrum, consistent with the spectral dependence of the CIB at the other simulated frequencies.
\begin{figure*}
     \centering
     \begin{subfigure}[b]{0.45\textwidth}
         \centering
         \includegraphics[height=15cm,width=6cm]{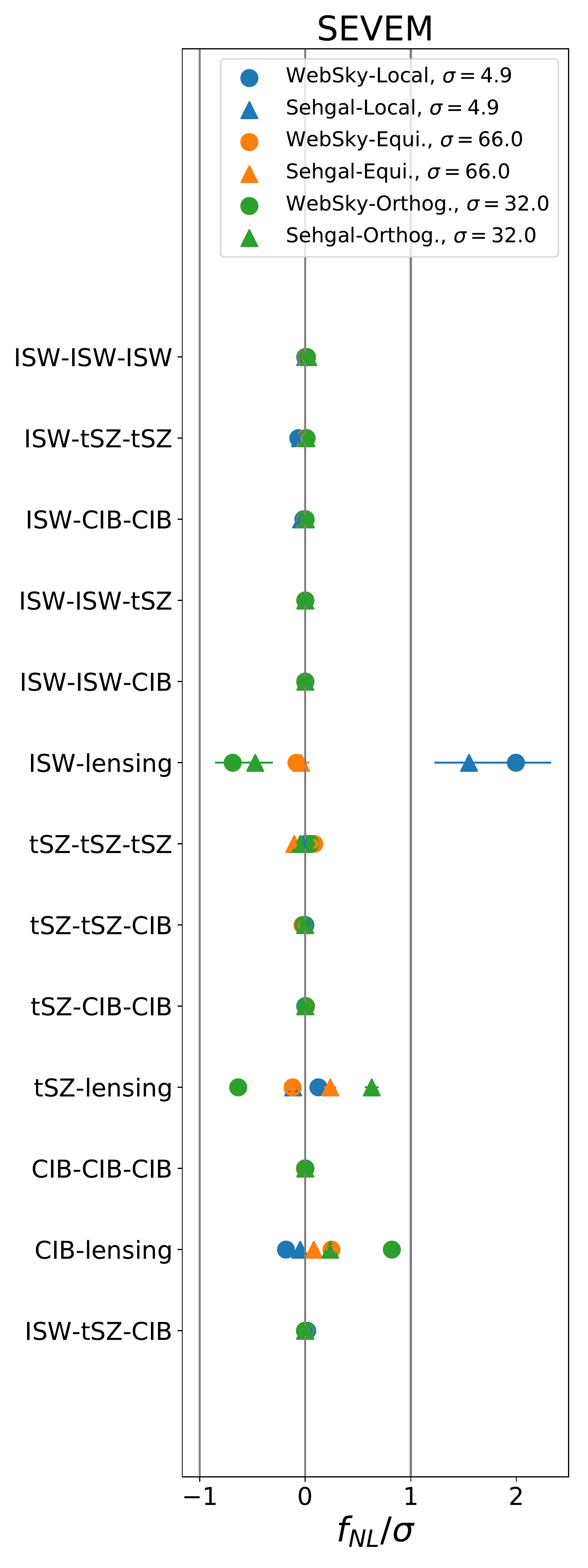}
         \caption{SEVEM}
         \label{fig:sevem_biases}
     \end{subfigure}
     \hfill
     \begin{subfigure}[b]{0.45\textwidth}
         \centering
         \includegraphics[height=15cm,width=6cm]{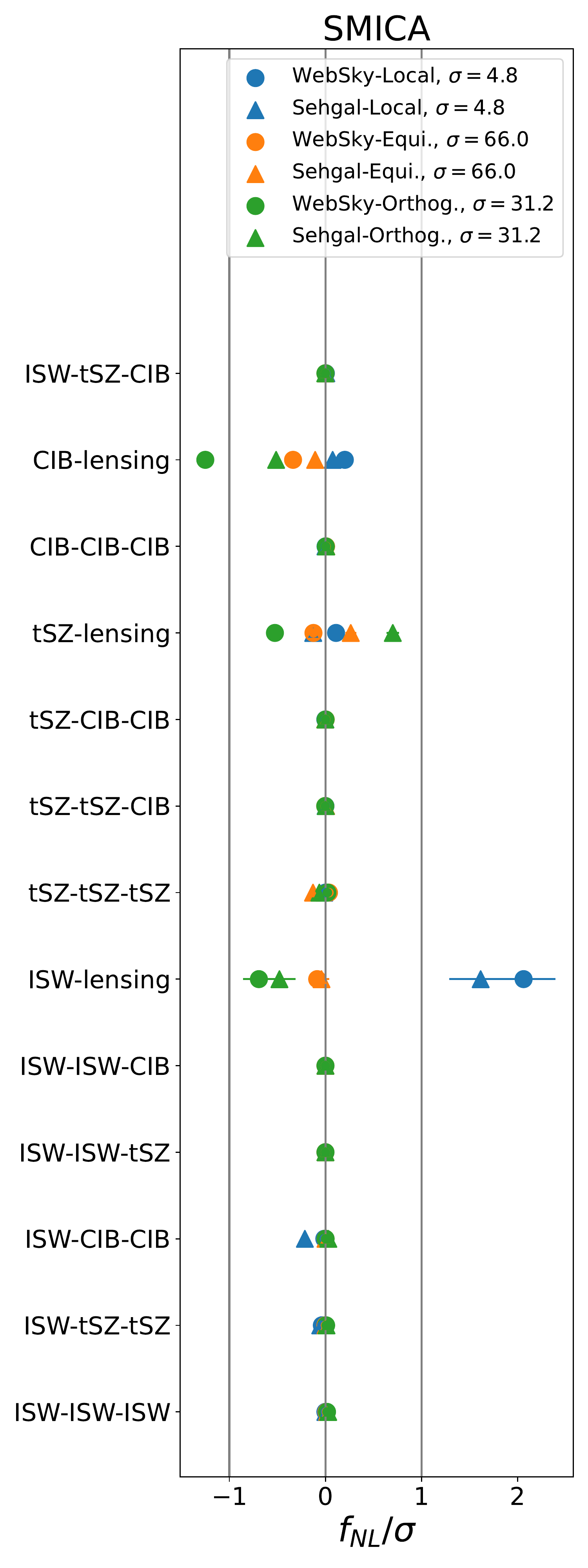}
         \caption{SMICA}
         \label{fig:smica_biases}
     \end{subfigure}
        \caption{Residual biases to a \textit{Planck}-like experiment with $\ell_{\textrm{max}}=1940$ for two component-separation methods: the template based SEVEM method and the parametric SMICA method. Otherwise, the specification is the same as Fig.~\ref{fig:Planck_ILC_biases}.}
        \label{fig:three graphs}
\end{figure*}

\subsection{SEVEM}
\subsubsection{SEVEM method}
The SEVEM method is primarily a pixel-based method; we briefly review the method here and refer the reader to \citet{leach2008,fernandez2012,planck2014-a11} for more details. It works by first constructing templates of foregrounds at different frequencies by differencing pairs of observations that have been smoothed to a common resolution (e.g., estimating dust and CIB by differencing the 353\,GHz and 545\,GHz maps). These foreground template maps can then be subtracted from the individual frequency maps to estimate cleaned maps at each frequency. The final step is to combine the cleaned frequency maps to construct a cleaned CMB map; this step is done in harmonic space with a set of $\ell$-dependent weights. The weights used in constructing the templates, cleaning frequency maps and producing the final cleaned CMB map are obtained by minimizing the variance and the residual foreground contributions. 

In this work we apply an approximate version of this pipeline where we follow all of the SEVEM steps except that we use weights that have been optimized for the Planck data, rather than our  simulations; these weights are given in \citet{planck2014-a11,planck2016-l04}. Given that the simulations are reasonably accurate representations of the data we expect that this will still provide a useful estimate of the level of residual biases.

\begin{figure}
\includegraphics[height=12cm,width=8cm]{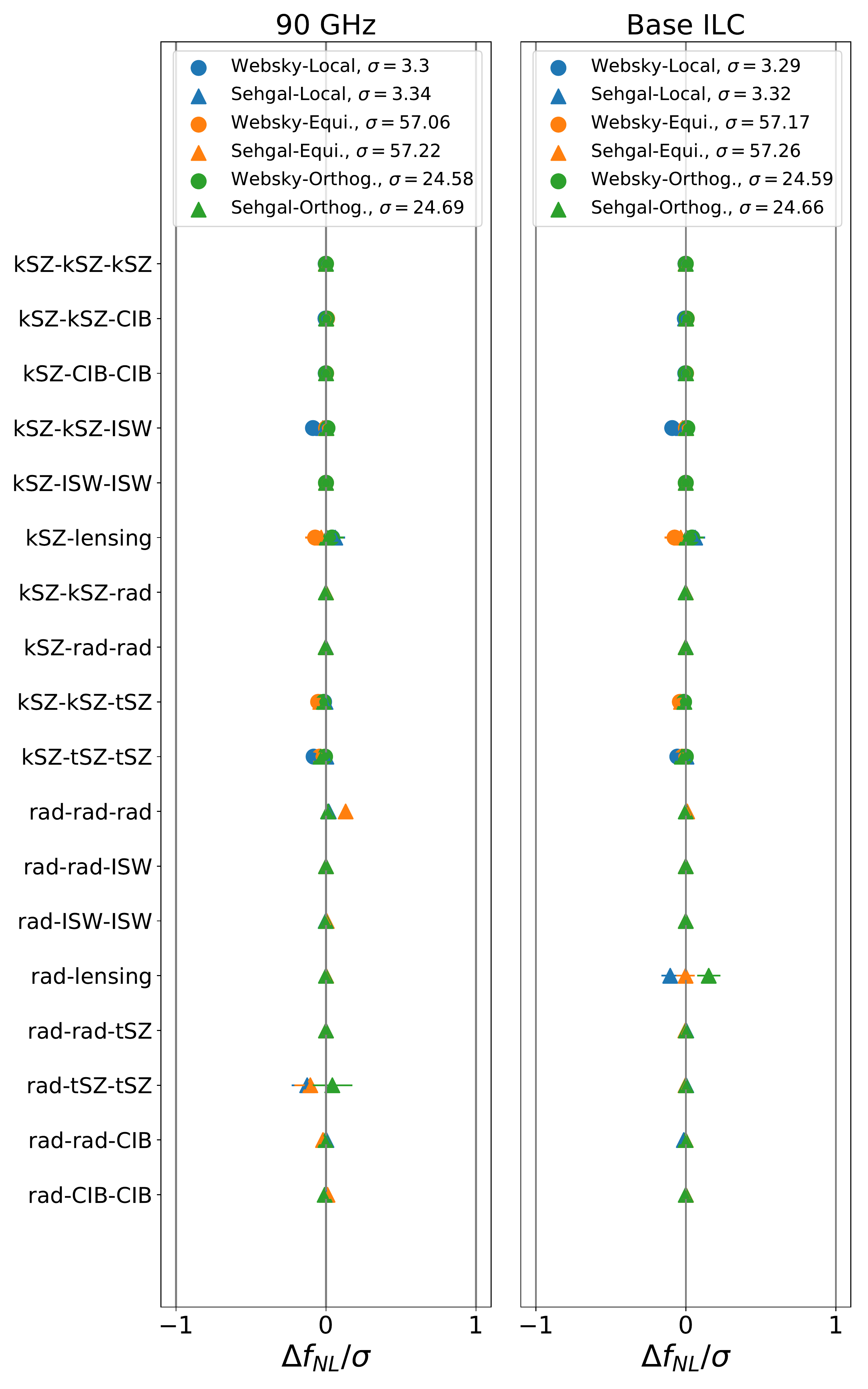}
\caption{Biases from the kSZ effect and from radio galaxies at 90\, GHz (left) for a SO-like experiment, and after a standard ILC applied to the combination of SO-like and \textit{Planck}-like experiments (right). Biases are normalised by the experiment errors (displayed in the legends of each panel). All biases are well below the experimental error and can be ignored in any non-Gauassianity analysis.}
\label{fig:rad+kSZ_Planck_sin_freq}
\end{figure}
\begin{figure}
\includegraphics[height=12cm,width=8cm]{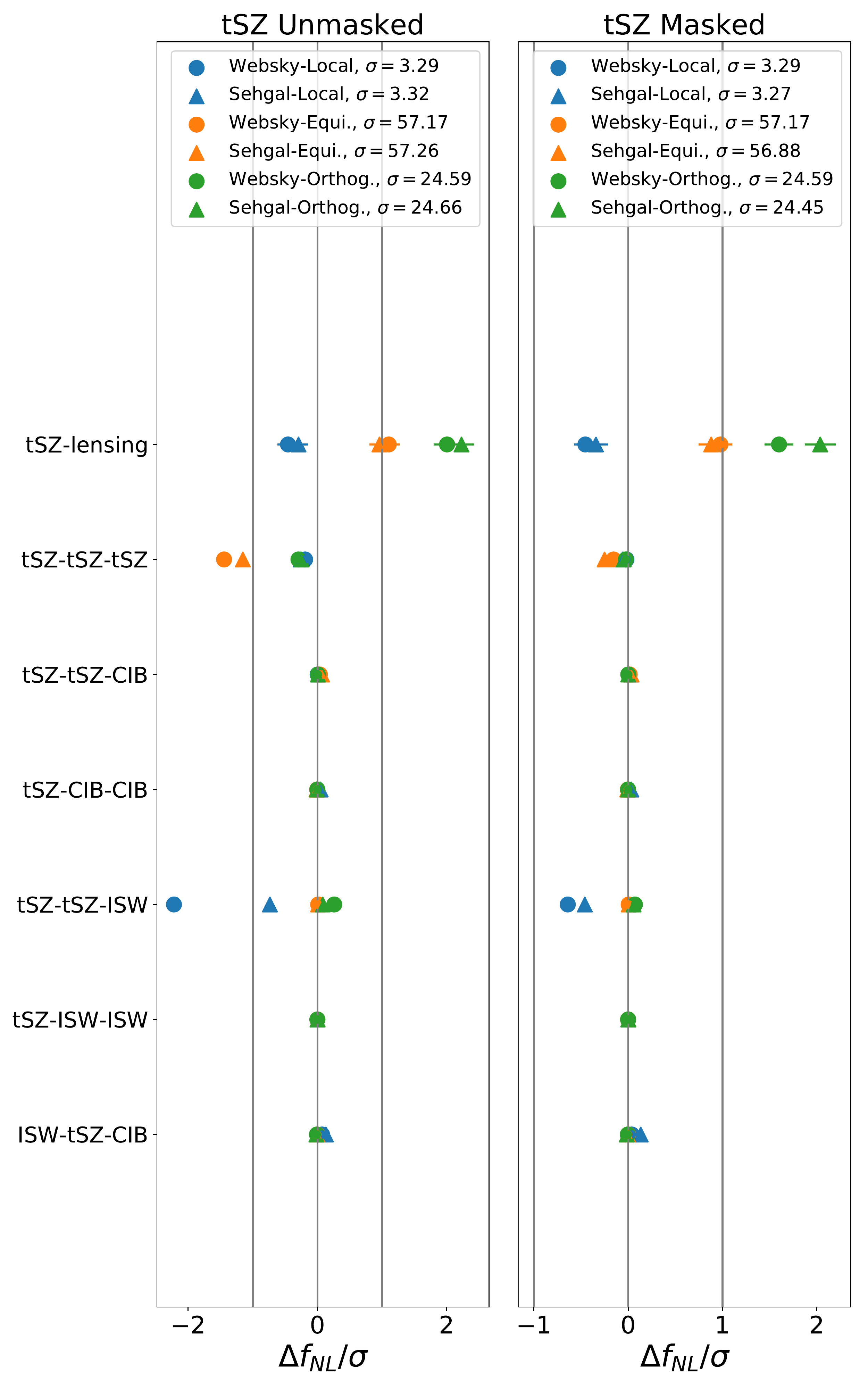}
\caption{Comparison of tSZ-related biases in the component-separated maps for an SO-like and \textit{Planck}-like combination before (left) and after (right) masking galaxy clusters. The most significant reductions are for the tSZ-tSZ-tSZ and tSZ-tSZ-ISW bispectra, with a larger reduction seen for the WebSky maps.}
\label{fig:tSZ_maskcomp}
\end{figure}
\subsubsection{Residual biases}
In Fig.~\ref{fig:sevem_biases}, we plot the biases found in the analysis of the SEVEM maps. We find that only lensing-related biases are non-negligible: the CIB-lensing, ISW-lensing and tSZ-lensing biases. 
Both the tSZ-lensing and CIB-lensing biases are less than the total experimental error. Thus we expect no significant biases in PNG analyses of the SEVEM maps.  Note that we see more significant differences between the Websky and Sehgal et al.~simulations here than was the case for the ILC used in the main text. This is likely due to the lower CIB power seen in the Sehgal et al.~simulations at high frequencies, compared to both Websky and the \textit{Planck} data \citep{Mak_2017}. 

\subsection{SMICA}
\subsubsection{SMICA method}
Here we briefly summarize the SMICA method and refer the reader to \citet{cardoso2008} for more details. The SMICA method is similar to the harmonic ILC method in that it constructs a clean map, $s_{\ell m}$, by a linear sum of the input maps, $\hat{d}_{\ell m}(\nu_i)$, in harmonic space:
\begin{align}
    \hat{s}_{\ell m} = \sum\limits_i w_{\ell m}(\nu_i)\hat{d}_{\ell m}(\nu_i).
\end{align}
The weights, $w_{\ell m}(\nu_i)$, are the same as the ILC (equation~\ref{eq:ILC})
\begin{align} 
    \hat{s}_{\ell m} = \sum\limits_{i j}\frac{a({\nu_i})[C_\ell^{-1}]^{\nu_i,\nu_j}\hat{d}_{\ell m}(\nu_j) }{a(\nu_i)[C_\ell^{-1}]^{\nu_i,\nu_j} a(\nu_j)}.
\end{align} 
Unlike the ILC method, which uses the empirical covariance matrix for $C_\ell^{\nu_i \nu_j}$, a model for the covariance matrix is used. The covariance matrix is modeled as the sum of three components: the CMB, foregrounds and noise. The power spectra and frequency dependence of the three components are chosen to be highly flexible functions with free parameters that are fit to match the empirical covariance matrix. To account for the strong spatial variation in the strength of Galactic foregrounds, the sky is decomposed into two parts: one centered on the Galaxy; and the other at high latitudes. These two regions are fit separately. 

Similar to the SEVEM method, we apply the SMICA method with weights fixed to the \textit{Planck} data rather than our simulations. This is done due to to complexity of refitting the weights.Note that the \textit{Planck} pipeline considered a second set of SMICA weights where  the tSZ signal was constrained (in a similar fashion to our tSZ-constrained ILC of Section \ref{sec:CleanedResults}). We elect here to only present the results obtained with the base SMICA weights. The second modification used in this work is that we only use the high-latitude weights. Using different weights over different parts of sky can introduce statistical anisotropy which leads to large increases in the computational complexity of bispectrum analyses. To avoid this computational cost we use just one set of weights. The high-latitude weights should be better at suppressing extragalactic components than the Galactic weights, as they are fit on a region of sky where the Galactic signals are significantly weaker, and thus are the appropriate weights to use for analysis of extragalactic biases. We verified that we obtain qualitatively similar results if we instead use the Galactic weights. 
\subsubsection{Residual Biases}
In Fig.~\ref{fig:smica_biases} we display the results from an analysis of maps that have been processed by our SMICA pipeline. We see that all of the potential biases are negligible with the exception of ISW-lensing, tSZ-lensing and CIB-lensing. The size of the CIB lensing bispectrum is sensitive to the modelling of the CIB and could contribute a bias at the $(0.5$--$1)\,\sigma$ level. The tSZ lensing bias is generally less than $0.5\, \sigma$. As in the SEVEM case, we see similar differences between the Websky and Sehgal et al.~simulations, likely arising from the different CIB spectral properties.

\section{Radio sources and kinetic Sunyaev--Zel'dovich results}\label{app:RadioAndKSZ}
We find that the biases from the radio sources and kinetic Sunyaev--Zel'dovich effects are negligible, both for single-frequency and foreground-cleaned observations. This can be seen in Fig.~\ref{fig:rad+kSZ_Planck_sin_freq}, where we show the biases to a single-frequency SO-like channel at 90\,GHz as well as the biases after component separation for combined observations from \textit{Planck}-like and SO-like experiments with $\ell_{\textrm{max}}=2960$. We see that all the biases are significantly below the experimental error. This is fortunate as removing kSZ-related biases would be challenging as the kSZ signal has the same frequency dependence as the CMB anisotropies.

\section{Impact of masking}
\subsection{ Point source masking}\label{app:PSmasking}
Bright point sources can cause challenges for component-separation methods and thus point sources are generally subtracted or inpainted before and after the application of component-separation methods \citep[see, e.g.,][]{planck2014-a11,planck2016-l04}. We found that this was similarly the case for our analysis. When applying the harmonic ILC to maps with bright CIB point sources, we often found large residuals and/or ringing patterns around these sources. It is thought that these effects arise due to the variation in the observed spectral energy distribution (SED) of the CIB sources: because the CIB has contributions from a broad redshift range, even if the rest-frame SEDs are identical the observed SEDs can have large variations and this leads to imperfect removal of point sources in the ILC. The large point-source residuals in the CIB maps lead to large non-Gaussianities in the maps and thus large variances on our estimates of the biases.

 A demonstration of this effect was found in our measurements of the CIB-CIB-CMB bispectrum. This bispectrum should be consistent with zero due to the unpaired primary CMB leg. However, when bright point sources were not masked, significant deviations from zero were observed. These arise due to large non-Gaussian contributions to the bispectrum variance, which are not captured by our simple Gaussian model. This effect is more prevalent in the Webky simulations where there is a much brighter tail of CIB sources.
 
The large non-Gaussianity could be modelled as the contribution from a small number of very bright Poisson-distributed sources. However, it is more optimal to remove it by masking.

To avoid this large variance we instead chose to mask these sources, akin to point-source masking using in the \textit{Planck} analyses. We constructed our point-source mask by identifying all sources with flux density brighter than 15\,mJy at 217\,GHz. This mask was then applied to the CIB maps at all frequencies. We applied this mask to the input CIB maps, which have no instrumental effects such as a beam applied. Only a very small number of pixels are masked, as each source only occupies a single pixel. We inpainted these masked pixels by assigning them the average of their neighbours. This masking procedure removes many more pixels from the Websky maps than the Sehgal et al.~maps.

This level of masking is appropriate for observations with SO but is overly aggressive for \textit{Planck}, where sources below 150\,mJy are not detectable.
As these bright sources are very rare we do not expect their removal to impact our results significantly. An alternative approach, following that usually applied to the data, would be to apply a point-source finder to simulated maps both before and after the ILC and explicitly remove the detected sources. This is beyond the scope of this work.

\subsection{Cluster masking}\label{app:clusterMasking}

Massive galaxy clusters will also be detectable at high significance in upcoming CMB experiments  via the tSZ effect \citep{SO_2019}. Whilst the tSZ spectral signature means the impact of these clusters can be mitigated through multi-frequency cleaning, as in Section~\ref{sec:CleanedResults}, we can exploit the highly non-Gaussian nature of this effect to mitigate these signals more effectively.

Such mitigation can be achieved by masking the brightest clusters. As the tSZ effect can be used to detect clusters almost independently of redshift, we choose a simple mass cut and mask all clusters with $M\geq 5 \times 10^{14} \,M_{\sun}h^{-1}$. The efficacy of this masking can be seen in Fig.~\ref{fig:tSZ_maskcomp}, where we see that cluster masking can reduce tSZ-related foreground biases in the component-separated maps by a factor of a few and is most effective for biases that are primarily sourced by the most massive clusters, e.g., the tSZ-tSZ-tSZ bispectrum\citep{Bhattacharya_2012}.

\bibliographystyle{mnras}
\bibliography{ngBiases,planck_bib}

\end{document}

%% file: Planck.tex
\def\setsymbol#1#2{\expandafter\def\csname #1\endcsname{#2}}
\def\getsymbol#1{\csname #1\endcsname}

\newbox\tablebox    \newdimen\tablewidth
\def\leaderfil{\leaders\hbox to 5pt{\hss.\hss}\hfil}
\def\tablenote#1 #2\par{\begingroup \parindent=0.8em
    \abovedisplayshortskip=0pt\belowdisplayshortskip=0pt
    \noindent
    $$\hss\vbox{\hsize\tablewidth \hangindent=\parindent \hangafter=1 \noindent
    \hbox to \parindent{$^#1$\hss}\strut#2\strut\par}\hss$$
    \endgroup}

\def\L2{\ifmmode L_2\else $L_2$\fi}

\def\DeltaT{\ifmmode \Delta T\else $\Delta T$\fi}
\def\deltat{\ifmmode \Delta t\else $\Delta t$\fi}
\def\fknee{\ifmmode f_{\rm knee}\else $f_{\rm knee}$\fi}
\def\Fmax{\ifmmode F_{\rm max}\else $F_{\rm max}$\fi}
\def\solar{\ifmmode{\rm M}_{\mathord\odot}\else${\rm M}_{\mathord\odot}$\fi}
\def\Msolar{\ifmmode{\rm M}_{\mathord\odot}\else${\rm M}_{\mathord\odot}$\fi}
\def\Lsolar{\ifmmode{\rm L}_{\mathord\odot}\else${\rm L}_{\mathord\odot}$\fi}
\def\inv{\ifmmode^{-1}\else$^{-1}$\fi}
\def\mo{\ifmmode^{-1}\else$^{-1}$\fi}
\def\sup#1{\ifmmode ^{\rm #1}\else $^{\rm #1}$\fi}
\def\expo#1{\ifmmode \times 10^{#1}\else $\times 10^{#1}$\fi}
\def\,{\thinspace}
\def\lsim{\mathrel{\raise .4ex\hbox{\rlap{$<$}\lower 1.2ex\hbox{$\sim$}}}}
\def\gsim{\mathrel{\raise .4ex\hbox{\rlap{$>$}\lower 1.2ex\hbox{$\sim$}}}}

\def\simprop{\mathrel{\raise .4ex\hbox{\rlap{$\propto$}\lower 1.2ex\hbox{$\sim$}}}}
\def\deg{\ifmmode^\circ\else$^\circ$\fi}
\def\pdeg{\ifmmode $\setbox0=\hbox{$^{\circ}$}\rlap{\hskip.11\wd0 .}$^{\circ}
          \else \setbox0=\hbox{$^{\circ}$}\rlap{\hskip.11\wd0 .}$^{\circ}$\fi}
\def\arcs{\ifmmode {^{\scriptstyle\prime\prime}}
          \else $^{\scriptstyle\prime\prime}$\fi}
\def\arcm{\ifmmode {^{\scriptstyle\prime}}
          \else $^{\scriptstyle\prime}$\fi}
\newdimen\sa  \newdimen\sb
\def\parcs{\sa=.07em \sb=.03em
     \ifmmode \hbox{\rlap{.}}^{\scriptstyle\prime\kern -\sb\prime}\hbox{\kern -\sa}
     \else \rlap{.}$^{\scriptstyle\prime\kern -\sb\prime}$\kern -\sa\fi}
\def\parcm{\sa=.08em \sb=.03em
     \ifmmode \hbox{\rlap{.}\kern\sa}^{\scriptstyle\prime}\hbox{\kern-\sb}
     \else \rlap{.}\kern\sa$^{\scriptstyle\prime}$\kern-\sb\fi}
\def\ra[#1 #2 #3.#4]{#1\sup{h}#2\sup{m}#3\sup{s}\llap.#4}
\def\dec[#1 #2 #3.#4]{#1\deg#2\arcm#3\arcs\llap.#4}
\def\deco[#1 #2 #3]{#1\deg#2\arcm#3\arcs}
\def\rra[#1 #2]{#1\sup{h}#2\sup{m}}

\def\dots{\relax\ifmmode \ldots\else $\ldots$\fi}
\def\WHzsr{\ifmmode $W\,Hz\mo\,sr\mo$\else W\,Hz\mo\,sr\mo\fi}
\def\mHz{\ifmmode $\,mHz$\else \,mHz\fi}
\def\GHz{\ifmmode $\,GHz$\else \,GHz\fi}
\def\mKs{\ifmmode $\,mK\,s$^{1/2}\else \,mK\,s$^{1/2}$\fi}
\def\muKs{\ifmmode \,\mu$K\,s$^{1/2}\else \,$\mu$K\,s$^{1/2}$\fi}
\def\muKRJs{\ifmmode \,\mu$K$_{\rm RJ}$\,s$^{1/2}\else \,$\mu$K$_{\rm RJ}$\,s$^{1/2}$\fi}
\def\muKHz{\ifmmode \,\mu$K\,Hz$^{-1/2}\else \,$\mu$K\,Hz$^{-1/2}$\fi}
\def\MJysr{\ifmmode \,$MJy\,sr\mo$\else \,MJy\,sr\mo\fi}
\def\MJysrmK{\ifmmode \,$MJy\,sr\mo$\,mK$_{\rm CMB}\mo\else \,MJy\,sr\mo\,mK$_{\rm CMB}\mo$\fi}
\def\microns{\ifmmode \,\mu$m$\else \,$\mu$m\fi}

\def\muK{\ifmmode \,\mu$K$\else \,$\mu$\hbox{K}\fi}
\def\microK{\ifmmode \,\mu$K$\else \,$\mu$\hbox{K}\fi}
\def\muW{\ifmmode \,\mu$W$\else \,$\mu$\hbox{W}\fi}
\def\kms{\ifmmode $\,km\,s$^{-1}\else \,km\,s$^{-1}$\fi}
\def\kmsMpc{\ifmmode $\,\kms\,Mpc\mo$\else \,\kms\,Mpc\mo\fi}
\providecommand{\sorthelp}[1]{}